\documentclass{kasper}

\usepackage{bm}
\usepackage{slashed}
\usepackage[bottom]{footmisc}
\newcommand{\norder}[2]{:\!\!#1\!\!:\!\!#2\,}

\DeclareMathOperator{\Tr}{Tr}

\DeclareMathOperator{\Imag}{Im}
\numberwithin{equation}{section}

\newdimen\tableauside\tableauside=1ex   %1.0ex
\newdimen\tableaurule\tableaurule=.32pt   %0.4pt
\newdimen\tableaustep
\def\phantomhrule#1{\hbox{\vbox to0pt{\hrule height\tableaurule width#1\vss}}}
\def\phantomvrule#1{\vbox{\hbox to0pt{\vrule width\tableaurule height#1\hss}}}
\def\sqr{\vbox{%
  \phantomhrule\tableaustep
  \hbox{\phantomvrule\tableaustep\kern\tableaustep\phantomvrule\tableaustep}%
  \hbox{\vbox{\phantomhrule\tableauside}\kern-\tableaurule}}}
\def\squares#1{\hbox{\count0=#1\noindent\loop\sqr
  \advance\count0 by-1 \ifnum\count0>0\repeat}}
\def\tableau#1{\vcenter{\offinterlineskip
  \tableaustep=\tableauside\advance\tableaustep by-\tableaurule
  \kern\normallineskip\hbox
    {\kern\normallineskip\vbox
      {\gettableau#1 0 }%
     \kern\normallineskip\kern\tableaurule}%
  \kern\normallineskip\kern\tableaurule}}
\def\gettableau#1 {\ifnum#1=0\let\next=\null\else
  \squares{#1}\let\next=\gettableau\fi\next}

\newcommand{\tabstack}[2]{\genfrac{}{}{0pt}{}{\vphantom{\tableau{1 1 1
1 1}}{#1}}{\mathbf{\scriptstyle #2}}}
\newcommand{\tabstackmed}[2]{\genfrac{}{}{0pt}{}{\vphantom{\tableau{1 1 1
1 1 1}}{#1}}{\mathbf{\scriptstyle #2}}}
\newcommand{\tabstackhigh}[2]{\genfrac{}{}{0pt}{}{\vphantom{\tableau{1 1 1
1 1 1 1}}{#1}}{\mathbf{\scriptstyle #2}}}
\newcommand{\oplusr}{\raise2ex\hbox{$\oplus$}}
\newcommand{\otimesr}{\raise2ex\hbox{$\otimes$}}
\newcommand{\lefttabbrack}[1]{\raise1ex\hbox{$\left#1\vphantom{\tableau{1 1 1 1 1 1 1 1 1 1}}\right.$}}
\newcommand{\righttabbrack}[1]{\raise1ex\hbox{$\left.\vphantom{\tableau{1 1 1 1 1 1 1 1 1 1}}\right#1$}}

\begin{document}
\title{The Ramond-Ramond sector of string theory beyond leading order}

\preprintnumber{AEI-2003-064\\CERN-TH/2003-177\\hep-th/0307298} 
\author{Kasper Peeters$^{1}$ and Anders Westerberg$^{2}$} 
\address{1}{{\it MPI/AEI f\"ur Gravitationsphysik}\\ 
{\it Am M\"uhlenberg 1}\\ 
{\it 14476 Golm,
Germany}} 
\address{2}{{\it CERN}\\
{\it TH-division}\\
{\it 1211 Geneva 23, Switzerland}}
\date{July 31st, 2003}
\email{kasper.peeters@aei.mpg.de, anders.westerberg@cern.ch}
\maketitle

\begin{abstract}
Present knowledge of higher-derivative terms in string effective
actions is, with a few exceptions, restricted to the NS-NS sector, a
situation which prevents the development of a variety of interesting
applications for which the RR~terms are relevant. We here provide the 
formalism as well as efficient techniques to determine the latter 
directly from string-amplitude calculations. As an illustration of these 
methods, we compute the dependence of the type-IIB action on the three- 
and five-form RR field strengths at four-point, genus-one, 
order-$(\alpha')^3$ level. We explicitly verify that our results are in 
accord with the SL(2,$\mathbb{Z}$) S-duality invariance of type-IIB string 
theory. Extensions of our method to other bosonic terms in the type-II 
effective actions are discussed as well.
\end{abstract}
\maketoc
\begin{sectionunit}
\title{Introduction}
\maketitle

A considerable amount of information about string and M-theory can be
extracted from the low-energy effective field theory actions, in
particular once one includes corrections that go beyond the leading
supergravity terms. A clear illustration of this fact can be found in
the large body of literature in which the effects of terms of higher
order in the Riemann tensor have been studied. This particular set of
corrections already provides an important testing ground for candidate
microscopic versions of \mbox{M-theory}~\cite{Helling:1999js}.  Upon
reduction to four dimensions, such higher-order terms influence the
couplings of the scalar
fields~\cite{Strominger:1998eb,Antoniadis:2003sw}.  They have also
been argued to lead to induced Einstein-Hilbert terms, which are
important for gravity localisation
phenomena~\cite{Antoniadis:2002tr,Antoniadis:1997eg} as they arise in
brane-world scenarios. Furthermore, they may play a role in
supersymmetry breaking~\cite{Becker:2002nn,Berg:2002es} by producing a
potential that stabilises certain moduli for non-supersymmetric vacua.
In black-hole and black-brane physics, higher-derivative terms lead to
modifications of the thermodynamics, and thus implicitly to new tests
of the string-theory description of the entropy of such
objects~\cite{Iyer:1994ys,deHaro:2003zd}.  Another important set of
applications can be found in the context of the AdS/CFT
correspondence, where the effects of higher-derivative terms in the
supergravity action map to~$1/N$ effects on the Yang-Mills
side~\cite{Gubser:1998nz,Pawelczyk:1998pb,Frolov:2001xr}. This list is
far from exhaustive, but it should suffice to illustrate the importance 
of having a good understanding of higher-derivative terms.

While higher-derivative terms are thus of considerable interest, 
they are very hard to obtain explicitly and this has been a
serious obstacle to further development of the applications just
mentioned. One particularly limiting factor is the fact that while
pure graviton corrections are relatively easy to compute, this is not
the case for terms involving any of the other supergravity fields. 
While the dependence on fermionic fields is perhaps not particularly
relevant except for supersymmetry considerations, there is a clear
need to determine the way in which the other bosonic fields appear
in the action. With~\cite{Peeters:2000qj} we have initiated a
programme to determine these terms using a combination of techniques
from string theory and supersymmetry. We refer the reader to the
introduction and discussion of that paper, as well as the
review~\cite{Peeters:2000sr}, for an overview of the existing
knowledge of higher-derivative terms in string and M-theory effective
actions and the methods used to obtain them.

The present paper is part of this long-term programme and aims at the
construction of certain terms that contain not just the graviton but
also gauge fields present in the supergravity multiplet.\footnote{We
should add at this point that an additional reason for being
interested in gauge-field terms is the role they play in modifying the
structure of supersymmetry transformations. In the context of
M-theory, we have argued in~\cite{Peeters:2000qj} that modifications
to the superspace torsion constraint in eleven dimensions are only
visible when the four-form gauge-field strength is taken into
account. A similar conclusion is expected in the ten-dimensional
theories. However, progress along these lines will also require the
construction of gauge-field terms involving fermions, something we
will not attempt in the present paper.}  Since we restrict ourselves
to bosonic action terms, a supersymmetry approach as used
in~\cite{Peeters:2000sr} would be very cumbersome, as the terms
involving fermions greatly outnumber the purely bosonic ones. Because
of the presence of spin-fields, a sigma-model $\beta$-function
calculation does not appear to be feasible either.\footnote{While the
approach of~\dcite{Berkovits:2001ue} allows one to deduce target-space
equations of motion in RR~backgrounds by solving BRST-nilpotency and
holomorphicity conditions, it seems rather complicated to extend their
results to three-loop level, which would be required to see
higher-derivative interactions in type-II theories.} For the
ten-dimensional theories, it is more straightforward to determine the
effective action directly from string-amplitude calculations, which is
what we will do in the present paper.  Our main aim is to present the
formalism and machinery behind the calculations that lead to a
determination of the terms involving gravitons and Ramond-Ramond gauge
fields (a topic which, rather surprisingly, has remained practically
untouched in the literature so far). In a certain sense our
calculations thus complete the work of~\dcite{Gross:1987mw}, which
only deals with the Neveu-Schwarz three-form field strength.
\medskip

In more detail, our calculations will be concerned with type-IIB
four-point functions at genus one, involving external three-form or
five-form Ramond-Ramond states. The restriction to four-point
functions has been made because many (though not all) conceptual and
technical problems are already present here; we will comment on
extensions to higher-point functions in the discussion.  The
restriction to genus one is motivated by the fact that at this order
in the string coupling, amplitudes with fewer than four external
particles vanish identically. Genus-one amplitudes thus only begin to
contribute to the effective action at the level of the~$(\alpha')^3
R^4$ term and the terms related to it by supersymmetry.  Furthermore,
genus-one calculations turn out to be technically simpler than their
cousins at genus-zero, and for the type-IIB string they are in any
case related by the SL(2,$\mathbb{Z}$) duality symmetry of the
effective field theory.

The first issue that requires attention is the calculation of the
world-sheet correlators. While closed-form expressions exist for
correlators of both the bosonic~\cite{DiVecchia:1989cy} and the fermionic
world-sheet fields~\cite{Atick:1987rs} for genus one (and in fact for
higher genera as well), the actual evaluation of our amplitudes is
still rather involved. As in previous calculations involving four
external particles, we eventually find that intricate cancellations
due to Riemann identities cause the integral over the vertex-operator
insertion points to reduce to an integral over a constant.  This is a
particular consequence of the low number of external particles and is
not expected to hold once higher orders are considered.

Subsequently, for reasons that will be explained, one needs an
efficient way to identify and subtract terms from the amplitude which
vanish by the the Bianchi identities, and to organise the result in 
terms of the tensors that appear in the effective action.  The spin 
fields in the world-sheet correlators, coming from the vertex operators 
of the Ramond-Ramond states, imply that the amplitude is composed of 
traces over (long strings of) gamma matrices, whose complexity increases 
rapidly with the number of gamma matrices. We will show how the amplitude 
and the resulting expressions for the effective action can be organised 
and analysed efficiently by employing group-theoretical arguments. This 
is illustrated most convincingly by the compact form of the final result 
of the five-form effective action, given in equation~\eqn{e:finalresult} 
below. 

As we have pointed out in the previous paper in this
series~\cite{Peeters:2001ub}, when doing this kind of amplitude
calculations one frequently uncovers various ill-understood aspects of 
string perturbation theory. The present paper is no exception to this
empirical rule, and we will again encounter various technical issues
which have received very little attention in the literature so far. As
such, our paper paves the way for the complete order-$(\alpha')^3$
analysis of the effective action, to be presented elsewhere, and more
generally we hope that our explicit and systematic presentation of the
calculation will be of practical use to others.
\medskip

As announced, our main result is the determination of the $R^2
(DF_{(5)})^2$ and $R^2 (DF_{(3)})^2$ terms in the effective action,
which are given in equations~\eqn{e:finalresult}
and~\eqn{e:finalresult3} respectively. Apart from the intrinsic
importance of actually being able to calculate these and similar other
terms completely and efficiently, we will use the last part of our
paper to emphasise that there are several reasons why these terms are
particularly interesting from a physical point of view.

Firstly, we will show how the three-form terms in the RR sector are
directly related, by SL(2,$\mathbb{Z}$) symmetry, to similar terms
involving the NS-NS three-form which were computed
by~\dcite{Gross:1987mw}. A strong check on our calculations is the
fact that our three-form action indeed precisely satisfies this
duality requirement, despite the fact that it arises in a completely
different way from the string calculations (the world-sheet
correlators in the \mbox{NS-NS} and \mbox{RR} sectors are completely
different). This provides a very interesting perturbative verification
of the SL(2,$\mathbb{Z}$) symmetry of type-IIB string theory.

Secondly, we will comment on the relation of our calculation to the
predictions of the linearised type-IIB superfield. It has recently
been shown by Berkovits and Howe that an extension of this superfield
construction to the non-linear level is problematic (a detailed
argument can be found in~\dcite{deHaro:2002vk}). This implies that our
string-based methods are at present the only way in which the
effective action terms at higher order in the fields can be reliably
determined. However, one still expects the superfield to be able to
predict four-field terms in the action. We will indeed show how the
superfield predictions at this level fit in precisely with our string
theory results.

We will end this paper with an outline of the computation of the $R^3
(F_{(5)})^2$ terms in the effective action, which are most interesting
from the point of view of physical applications and whose computation
will rely heavily on the results and methods we have derived here. We
will also comment on similar computations involving space-time fermions.
\end{sectionunit}

\begin{sectionunit}
\title{Four-boson string amplitudes}
\maketitle
\begin{sectionunit}
\title{World-sheet correlators and modular integrals}
\maketitle
\label{s:stringstuff}

When computing string amplitudes, there are in general many different
ways in which the momenta and polarisation tensors can be contracted
in the final result. Our approach will be to classify these
Lorentz-invariant contractions using a group-theoretical method; this
is the key element that allows us to present the amplitude, and later
on the resulting effective action, in a compact form. However, there are
many Lorentz-invariant combinations which are simply incompatible with
the exchange symmetries of the external states in the
amplitude. Before doing the classification, we would like to eliminate
such incompatible contractions.

Since we are interested in the $R^2 (DF_{(5)})^2$ terms in the effective 
action, we will have to compute the string amplitude with two gravitons
and two five-forms.  This amplitude is manifestly invariant under
the exchange of e.g.~the two five-form vertex operators, which
involves the exchange of the polarisation tensors~$F^{(1)}$
and~$F^{(2)}$, the momenta~$k^{(1)}$ and~$k^{(2)}$, as well as the
insertion point variables~$z_1$ and~$z_2$. In order to deduce the
symmetry properties under exchange of the polarisation tensors and
external momenta only, we need to analyse the $z$-dependence of
the string integrand. Understanding the $z$-dependence is also
important because we eventually wish to compute the full amplitude,
not just the tensorial structures, and we will therefore have to
perform the final integral over all the insertion point variables. In
the present section we determine this \mbox{$z$-dependence} of the
combined world-sheet correlators.

\begin{table}[t]
\begin{equation*}
\begin{aligned}
Y \quad && V_F^{(-1/2,-1/2)} \quad && V_F^{(-1/2,-1/2)}\quad && V_\zeta^{(0,0)} \quad && V_\zeta^{(0,0)} \quad\\[1ex]
\partial X \Psi  \quad         && S_{\text{L}}    \quad      && S_{\text{L}}       \quad   &&
\partial X + k\Psi\Psi \quad && \partial X + k\Psi\Psi \quad  \\[1ex]
                          && \Gamma^{[5]}e^{ik_1\cdot X}\quad &&
                          \Gamma^{[5]}e^{ik_2 \cdot X}\quad &&
e^{ik_3\cdot X}\quad  && e^{ik_4\cdot X} \quad 
                          \\[1ex]
\bar\partial X \tilde\Psi\quad && S_{\text{R}} \quad         && S_{\text{R}} \quad         &&
\bar\partial X + k\tilde\Psi\tilde\Psi\quad && \bar\partial X +
                          k\tilde\Psi\tilde\Psi\quad  \\[1ex]
\end{aligned}
\end{equation*}
\caption{Schematic form of the two-graviton, two-five-form string
  amplitude. We have indicated the ghost numbers of the vertex
  operators, and split the correlator in left- and right-moving
  parts. $Y$~denotes the picture changing operator.}
\label{t:correlator}
\end{table}
The correlator for the two-graviton, two-five-form amplitude can be
summarised using a notation we have used also
in~\cite{Peeters:2000qj}; see table~\ref{t:correlator}. This is based
on the explicit form of the five-form vertex operator in the
$(-\tfrac{1}{2},-\tfrac{1}{2})$~ghost picture,
\begin{equation}
\label{e:fiveform_vop}
V^{(-1/2,-1/2)}_{F_{(5)}}(k) = \frac{1}{5!}\int\!{\rm d}^2 z\, F_{mnpqr}\, 
  (\bar S_{\text{L}} \Gamma^{mnpqr} S_{\text{R}})\, e^{-\phi/2-\tilde\phi/2} e^{ik\cdot X}\, ,
\end{equation}
as well as on the standard one for the graviton with $(0,0)$~ghost
charges,
\begin{equation}
\label{e:graviton_vop}
V^{(0,0)}_{\zeta}(k) = \int\!{\rm d}^2 z\, \zeta_{mn} 
  (\partial X^m - i k\cdot \Psi \Psi^m)(\bar\partial X^n - i k\cdot\tilde\Psi\tilde\Psi^n)
  \,e^{ik\cdot X}\, .
\end{equation}
(We will comment below on the fact that these vertex operators should, in 
principle, still be amended with world-sheet gravitino terms.)  Using the 
fact that fermionic correlators of the form~$\langle S S\, \Psi\rangle$ 
vanish identically, there are thus, a~priori, two different types of terms 
in each chiral sector of the string amplitude: those with three RNS 
fermions~$\Psi$ and those with five. We will argue below that the terms
involving $\langle S S\, \Psi\!\norder{\Psi\Psi}{}\,\rangle$ correlators 
do not contribute at eight-derivative order in the amplitude, so that the 
full four-particle amplitude is given by the single term
\begin{multline}
\label{e:fourpointampl}
\mathcal{A} =  k^{(3)}_{r_1} k^{(3)}_{s_1} k^{(4)}_{r_3} k^{(4)}_{s_3} 
                \zeta^{(3)}_{r_2s_2} \zeta^{(4)}_{r_4s_4} 
        \int_{\cal F}\!\frac{{\rm d}^2\tau}{\Imag\tau}
        \prod_{i=1}^4 \int_{\cal T} \!{\rm d}^2 z_i\\[1ex]
     \lim_{\begin{aligned}[t] \scriptstyle v &\scriptstyle \rightarrow
           \scriptstyle z_1\\[-1ex]
                              \scriptstyle w &\scriptstyle \rightarrow
           \scriptstyle z_2 \end{aligned}}\bigg[
           \,\mathcal{B}_{{m} {n}}\,
           \Tr\Big( {\slashed{F}}^{(1)} \tilde{\mathcal{F}}^{{n} s_1s_2 s_3s_4}
                 ({\slashed{F}}^{(2)})^\mathrm{T} 
                 (\mathcal{F}^{{m} r_1r_2 r_3r_4})^\mathrm{T}
              \Big)
           \, {\cal G}\tilde{\cal G}\bigg]\, .
\end{multline}
The bosonic correlator that appears here is given by
\begin{align}
\label{e:fourpointbosons}
\mathcal{B}^{{m} {n}} &= \big\langle \partial X^{{m}}(v) \bar\partial
X^{{n}}(w) \prod_{i=1}^4 \exp\left[ i k_i\cdot X(z_i)\right] \big\rangle\,
,
\intertext{while the fermionic and ghost correlators are}
\label{e:fourpointfermions}
\mathcal{F}^{{m} r_1 r_2 r_3 r_4} &= \big\langle \Psi^{{m}}(v)\, S_{\text{L}}(z_1) S_{\text{L}}(z_2) 
\norder{\Psi^{r_1}\Psi^{r_2}}{(z_3)} 
\norder{\Psi^{r_3}\Psi^{r_4}}{(z_4)}\big\rangle\, ,\\[1ex]
\mathcal{G} &= \big\langle \exp\left[-\phi(v)\right]\,\exp\left[ -\phi(z_1)/2\right]
\,  \exp\left[-\phi(z_2)/2\right]\big\rangle\,,
\end{align}
with similar expressions for the right-moving sector. These can
all be computed explicitly: techniques for both the bosonic and the
fermionic correlators have been given by~\dcite{Atick:1987rs}.
Details can be found in tables~\ref{t:fourpointbosons}
and~\ref{t:twotwo}.

Since the holomorphic and anti-holomorphic fermionic correlators
come with a $(v-z_1)$ and a $(\bar w-\bar z_2)$ zero, respectively, the
only relevant term in the bosonic correlator given above is the
momentum-dependent one containing a $(v-z_1)^{-1}(\bar w-\bar
z_2)^{-1}$ pole. The terms coming from the zero-mode contraction of
the bosons in the two picture changers, for instance, do not
contribute to the amplitude.  Note also that it is, a priori,
possible to obtain second-order momentum factors from the correlator
of the exponentials. However, at genus one, this correlator of
exponentials leads to an expansion $1 + (\alpha')^3 k^6 + \cdots$,
i.e.~the order $k^2$ and $k^4$ terms are absent (this can be deduced
from the explicit calculation of the correlator of exponentials
by~\dcite{Green:1999pv}; see the discussion around equation (5.2) in
that paper).

\begin{table}[t]
\begin{equation*}
\mathcal{B}^{{m} {n}} = \Big[\begin{aligned}[t]
{}&\Big(\sum_{i=1}^4 k_i^{{m}} \big[ - i \partial_v \ln\theta_1(v-z_i) + \frac{2\pi}{\Imag{\tau}}
\Imag{(v-z_i)} \big]\Big)\\[1ex]
{}&\times \Big(\sum_{j=1}^4 k_j^{{n}} \big[ - i \bar\partial_w
\ln\bar\theta_1(\bar w-\bar z_j) - \frac{2\pi}{\Imag{\tau}}
\Imag{(w-z_j)} \big]\Big)\\[1ex]
&{}- \frac{2\pi}{\Imag\tau} \eta^{{m}{n}}\Big]\times \Big\langle \prod_{m=1}^4 e^{i k_m\cdot X(z_m)}\Big\rangle\, ,
\end{aligned}
\end{equation*}
\caption{Explicit expression for the bosonic correlator
$\langle \partial X \bar\partial X\, \prod_{i=1}^4 e^{ikX}\rangle$.
Despite the appearance of the sum, only one term survives when this
correlator is combined with the fermionic one. Note that our
way of taking care of the picture changer differs from the method
in~\dcite{Atick:1987rs} (who take the limit $v\rightarrow z_1$ before
computing the bosonic correlator); the result is, however, equivalent.}
\label{t:fourpointbosons}
\end{table}
\begin{table}[t]
\begin{center}
\begin{tabular}{llllllllrrrr}
structure & $r_1$ & $r_2$ & $r_3$ & $r_4$ & $m$ & $\alpha_1$ &
$\alpha_2$ & $\Gamma$ & $\langle\cdots\rangle/K_\nu$ & $K_\nu$ & coeff \\[1.5ex]
$(\Gamma^{r_1r_2r_3r_4m})_{\alpha_1\alpha_2}$             & 1 & 3 & 5 & 7 & 8 &
$\scriptstyle (---++)$ & $\scriptstyle (-----)$ & $-i$ & $-i(v-z_1)$ & 1 & 1\\[2ex]
$(\Gamma^{r_1r_2r_3})_{\alpha_1\alpha_2}\,\delta^{r_4m}$    & 1 & 3 & 5 & 7 & 7 & 
$\scriptstyle (---++)$ & $\scriptstyle (-----)$ & $\tfrac{1}{2}$ & $(v-z_1)$ & 1 & 2 \\[.5ex]
$(\Gamma^{r_3r_4r_1})_{\alpha_1\alpha_2}\,\delta^{r_2m}$    & 5 & 7 & 1 & 3 & 7 & 
$\scriptstyle (---++)$ & $\scriptstyle (-----)$ & $\tfrac{1}{2}$ & $(v-z_1)$ & 1 & 2\\[2ex]
$(\Gamma^{r_2r_4 m})_{\alpha_1\alpha_2}\,\delta^{r_1r_3}$  & 1 & 3 & 1 & 5 & 7 & 
$\scriptstyle (+---+)$ & $\scriptstyle (-----)$ &  & $0$ & & $0$ \\[2ex]
$(\Gamma^{r_1})_{\alpha_1\alpha_2}\,\delta^{r_2r_3} \delta^{r_4m}$ & 1 & 3 & 3 & 5 & 5 &
$\scriptstyle (-++++)$ & $\scriptstyle (-----)$ & $\tfrac{1}{4}$ & $-2(v-z_1)$ & $-1$ & $8$ \\[.5ex]
$(\Gamma^{r_3})_{\alpha_1\alpha_2}\,\delta^{r_4r_1} \delta^{r_2m}$ & 3 & 5 & 1 & 3 & 5 &
$\scriptstyle (-++++)$ & $\scriptstyle (-----)$ & $\tfrac{1}{4}$ & $-2(v-z_1)$ & $-1$ & $8$ \\[2ex]
$(\Gamma^{m})_{\alpha_1\alpha_2}\,\delta^{r_1r_3}\delta^{r_2r_4}$  & 1 & 3 & 1 & 3 & 5 &
$\scriptstyle (++-++)$ & $\scriptstyle (-----)$ & $1$ & $(v-z_1)$ & 2 & 2
\end{tabular}
\end{center}
\caption{Decomposition into irreducible tensor structures of the
fermionic correlator $\langle \Psi(v)\, S(z_1) S(z_2) \!
\norder{\Psi\Psi}{(z_3)}\!\norder{\Psi\Psi}{(z_4)}\rangle$. These
expressions should still be antisymmetrised, with unit weight, in
$r_1,r_2$ and $r_3,r_4$ respectively.  The seven structures correspond
to the number of singlets in the
$\mathbf{10}\otimes\mathbf{16}\otimes\mathbf{16}\otimes\mathbf{45}\otimes\mathbf{45}$
tensor product. We are using the picture changing choice $v\rightarrow
z_1$. The values of the $r$-indices refer to real world-sheet
fermions, which are related to the complex SO(2) fermions by $\Psi^1 =
\psi + \bar\psi$ and $\Psi^2 = -i (\psi - \bar\psi)$ and similarly for
the other eight.  The values of the spinor indices are chosen such
that a non-zero entry of the~$\Gamma$ matrix product is selected; this
number is listed in the ``$\Gamma$'' column (the explicit expressions
for these matrices in the particular basis that we are using can be
found in appendix~A.2 of~\cite{Peeters:2001ub}).  The last three
columns give the result of the ``unnormalised'' Atick\&Sen expressions
presented in appendix~\ref{a:fermcorr}, the normalisation factor and the
final result for the coefficient of each tensorial structure in the
amplitude (obtained when the pole of the bosonic correlator is taken
into account).}
\label{t:twotwo}
\end{table}

This balance between zeroes from the fermionic correlators and poles
from the bosonic ones is also the reason why the amplitude does not
receive any contributions from $\langle S S\,
\Psi\!\norder{\Psi\Psi}{}\,\rangle$ correlators: these can be shown to
be proportional to a double zero in each chiral half, given by
$(v-z_1)^2 (\bar{w}-\bar{z}_2)^2$. The corresponding bosonic
correlator does contain a second order pole, but only at order~$k^4$
in the external momenta (the resulting term in the amplitude
presumably talks to field-theory subtraction terms from the effective
action at lower order). For our calculation these terms are, in any
case, irrelevant.

Note that we are in the fortunate situation that our amplitude does
not receive any contributions from the pinch singularities discussed by,
e.g., \dcite{Minahan:1988ha} and~\dcite{Bern:1990fu}. That is, our
amplitude does not contain integrals of the form $\int\!{\rm
d}^2\nu_i\, k_i\cdot k_j|\nu_i-\nu_j|^{-2-k_i\cdot k_j}=1$. This is
simply because our bosonic poles do not survive but instead get
cancelled by the zeroes from the fermionic correlator.

We should at this stage also comment on the presence of terms in the
vertex operators which involve the world-sheet gravitinos. As we have
pointed out in~\cite{Peeters:2001ub}, these terms sometimes result in
additional contributions to the amplitude. In the present situation,
however, world-sheet gravitino terms are harmless. This is essentially
because the replacement of a picture changer with a term involving a
world-sheet gravitino results in a bosonic correlator in which both
a~$\partial X$ and a~$\bar\partial X$ factor have been removed (this
can be deduced simply from the BRST transformation rules as spelled
out in equation~(2.7) of~\cite{Peeters:2001ub}).  In the present
situation this means that the bosonic correlator reduces to the
correlator of plane waves, while the fermionic correlator remains
unchanged.  The zero of the latter is then no longer balanced by a
pole from the bosonic correlator, and as a result the contribution of
these world-sheet gravitino terms to the amplitude
vanishes.\footnote{This point can be summarised by the general rule of
thumb that the world-sheet gravitino terms in vertex operators are
\emph{only} relevant for amplitudes in which, before the replacement
of picture changers with world-sheet gravitinos, there are non-trivial
contributions arising from the zero-mode part of a~$\langle \partial X
\bar\partial X\rangle$ contraction.}
%
% Compare (3.15) and (3.18) of izletter.
%
%  dX Psi :Psi Psi:  ->  :PsiPsiPsi:
%           dbarX
%
%
% So one loses both dX and dbarX in this case, yet the fermionic
% correlator remains completely unchanged. 

Just as for the four-graviton amplitude, we observe that after the use
of the Riemann identity, and after the picture-changing limits have
been taken, the integral over the insertion points and the modular
parameter becomes trivial, as the integrand reduces to a
constant.\footnote{The Riemann identity~\eqn{e:Riemann_id} can be used 
  by virtue of the fact that one of the $\theta_\nu(\frac{z_1-z_2}{2})$ 
  factors of the fermionic correlator cancels against a part of the ghost 
  correlator,
\begin{equation}
{\cal G}(v,z_1,z_2) = 
\frac{\theta_1\big(v-z_1\big)^{\tfrac{1}{2}}
  \theta_1\big(v-z_2\big)^{\tfrac{1}{2}}}{\theta_1\big(z_2-z_1\big)^{\tfrac{1}{4}} \theta_\nu\big(v-\frac{z_1+z_2}{2}\big)}
\end{equation}
  after the limit $v\rightarrow z_1$ has been taken.} 
This implies that symmetry properties of the amplitude under exchange of 
vertex operators reduce simply to symmetry properties under exchange of
momenta and polarisation tensors (the exchange of the insertion points
becomes a trivial operation). As we will see in the next section, this 
fact significantly reduces the number of Lorentz-invariant structures
that need to be taken into account. With the basis of Lorentz invariants 
that we will construct there, we can then go back to the string amplitude 
and compute the final step, namely the trace of the gamma matrices
in eq.~\eqn{e:fourpointampl}. This calculation will be tackled in
section~\ref{s:amplituderesults}.
\end{sectionunit}

\begin{sectionunit}
\title{Group-theoretical classification of terms in the effective action}
\maketitle
\label{s:invclass}

Having determined the world-sheet correlators which form the building
blocks of the string amplitude, we are now in a position to determine
a general basis of four-field terms, compatible with that string
amplitude, in which we will express the effective action.  For certain
applications it may be useful to have access to the explicit index
contractions of the two Weyl tensors and two five-form fields
(together with their derivatives). However, here we will present our
results using a more compact notation, based on a group-theoretical
classification of the various factors that make up a term. More
details of such group-theoretical decompositions can be found in
\dcite{Fulling:1992vm}.

There are several reasons for adopting this approach. One is that the
it allows for a nice and compact way of expressing the effective
action which is completely general and does not rely on any a~priori
knowledge about the origin of the various terms.\footnote{For the
four-graviton terms, a compact notation is achieved by using the $t_8$ 
and $\epsilon_{10}$ symbols, by means of which the type-IIB four-graviton
interaction term can be written as
$(t_8t_8+\tfrac{1}{8}\epsilon_{10}\epsilon_{10})W^4$. However, these 
tensors do not suffice for the terms under consideration in the present 
paper. (See~\cite{Peeters:2000qj} for another set of terms, namely
fermi bilinears, where more complicated tensor structures arise.)} At
the practical level, this way of organising the result has also
provided very strong checks on signs and factors in the
calculation. Finally, and most importantly, it allows for a systematic 
identification of the terms proportional to Bianchi identities. This is 
crucial when analysing the amplitudes directly in terms of the RR field 
strengths that appear in the vertex operators, since the physical state
conditions for these fields correspond to the requirements
${*}d{*}F_{(5)}^+ = 0 = d F_{(5)}^+$ (with analogous expressions for
the RR three-form). In coordinate notation, the former condition just
translates to $D^k F_{km_2m_3m_4m_5}^+=0$ and is therefore trivial to
handle.  Imposing the latter, however, requires that the fully
antisymmetrised tensors $D^{\phantom{+}}_{[m}F_{m_1m_2m_3m_4m_5]}^+$ be identifiable
in general $W^2(DF_{(5)}^+)^2$ contractions, something which the
group-theoretical decompositions allow us to achieve.

For the $W^2 (D F_{(5)}^+)^2$ terms in the action, it is useful to
first group together terms according to the rank of the $W^2$ and
$(DF_{(5)}^+)^2$ blocks, i.e.~by the number of indices which contract
between the two blocks. These quadratic blocks can then be decomposed
further into distinct irreducible representations. The two basic
tensors sit in the representations
\begin{equation}
\raise2ex\hbox{$W:$} \tabstackmed{\widetilde{\tableau{2 2}}}{[02000]}\,\raise2ex\hbox{,}\quad\quad\quad
\raise2ex\hbox{$D F_{(5)}^+:$}
\tabstackmed{\tableau{1}}{[10000]} \otimesr \tabstackmed{\tableau{1 1 1 1 1}^+}{[00002]} \raise2ex\hbox{$=$}
\tabstackmed{\tableau{1 1 1 1 1 1}}{[00011]} \oplusr\tabstackmed{\widetilde{\tableau{2
1 1 1 1}}^+}{[10002]}\, \raise2ex\hbox{.}
\end{equation}

The symmetric product of two Weyl tensors yields the following
decomposition in terms of Young tableaux:
% 		 1X[0,0,0,0,0] +1X[0,0,0,1,1] +1X[0,0,0,2,2] +1X[0,0,2,0,0] +
% 				  0              4              8             6
% 
% 		 2X[0,2,0,0,0] +1X[0,2,0,1,1] +1X[0,4,0,0,0] +1X[1,1,1,0,0] +
% 				  4              8              8             6
% 
% 		 1X[2,0,0,0,0] +1X[2,0,0,1,1] +1X[2,0,2,0,0] +1X[2,2,0,0,0] +
% 				  2              6              8             6
% 
% 		 1X[4,0,0,0,0]
% 				  4
\begin{equation}
\label{e:W2reps}
\raise2ex\hbox{$\Big( \widetilde{\tableau{2 2}} \otimes \widetilde{\tableau{2 2}}\Big)_s =$}
\begin{aligned}[t]
 \tabstack{\widetilde{\tableau{2 2 2 2}}}{[00022]} &\oplusr
  \tabstack{\widetilde{\tableau{4 2 2}}}{[20200]} \oplusr
  \tabstack{\widetilde{\tableau{3 3 1 1}}}{[02011]} \oplusr
  \tabstack{\widetilde{\tableau{4 4}}}{[04000]} \\[1ex]
 &\oplusr\tabstack{\widetilde{\tableau{2 2 2}}}{[00200]} \oplusr
          \tabstack{\widetilde{\tableau{3 2 1}}}{[11100]} \oplusr
          \tabstack{\widetilde{\tableau{3 1 1 1}}}{[20011]} \oplusr
          \tabstack{\widetilde{\tableau{4 2}}}{[22000]} \\[1ex]
 &\oplusr\tabstack{\tableau{1 1 1 1}}{[00011]} \oplusr
			  \tabstack{2\,\widetilde{\tableau{2 2}}}{[02000]} \oplusr
			  \tabstack{\widetilde{\tableau{4}}}{[40000]} 
 \oplusr\tabstack{\widetilde{\tableau{2}}}{[20000]} \oplusr
			  \tabstack{\cdot}{[00000]}\, \raise2ex\hbox{.}
\end{aligned}
\end{equation}
However, not all of these representations are compatible with the
symmetries of the amplitude. For instance, the rank-eight contractions
are fully antisymmetric in the four indices of the ``$r$'' and
``$s$'' sets respectively, as can be deduced from the form of the
fermionic correlator given in table~\ref{t:twotwo}. As a consequence,
of the four a~priori possible rank-eight tableaux, only the $[00022]$
can appear in the amplitude. Similar arguments can be used to discard
the $[11100]$ and $[22000]$, leaving $[00200]$ and $[20011]$ as the
only rank-six tableaux. All tableaux of rank four and lower remain,
however.

The product of the two $DF_{(5)}^+$ tensors can be decomposed
similarly. Keeping only the tableaux that are compatible with the
symmetries of the string amplitude, one finds
\begin{subequations}
\label{e:DF5squareexpansions}
\begin{align}
\lefttabbrack(\tabstackmed{\widetilde{\tableau{2 1 1 1 1}}^+}{[10002]} \otimesr
\tabstackmed{\widetilde{\tableau{2 1 1 1 1}}^+}{[10002]} \righttabbrack)_{\!\!\!s}  
&\raise2ex\hbox{$\rightarrow$}
&\tabstack{\widetilde{\tableau{2 2 2 2}}}{[00022]}
&\oplusr \tabstack{\widetilde{\tableau{2 2 2}}}{[00200]}
&\oplusr \tabstack{\widetilde{\tableau{3 1 1 1}}}{[20011]}
&\phantom{\oplusr \tabstack{2\,\tableau{1 1 1 1}}{[00011]}} 
&\oplusr \tabstack{\widetilde{\tableau{2 2}}}{[02000]} 
&\oplusr \tabstack{\widetilde{\tableau{4}}}{[40000]} 
&\oplusr \tabstack{\widetilde{\tableau{2}}}{[20000]} 
&\phantom{\oplusr \tabstack{\cdot}{[00000]}}
\\[1ex]
\lefttabbrack(\tabstackmed{\widetilde{\tableau{2 1 1 1 1}}^+}{[10002]} \otimesr
\tabstackmed{\tableau{1 1 1 1 1 1}}{[00011]} \righttabbrack)  
&\raise2ex\hbox{$\rightarrow$}
&\tabstack{\widetilde{\tableau{2 2 2 2}}}{[00022]}
&\oplusr \tabstack{\widetilde{\tableau{2 2 2}}}{[00200]}
&\oplusr \tabstack{\widetilde{\tableau{3 1 1 1}}}{[20011]}
&\oplusr \tabstack{\tableau{1 1 1 1}}{[00011]} 
&\oplusr \tabstack{\widetilde{\tableau{2 2}}}{[02000]} 
&\phantom{\oplusr \tabstack{\widetilde{\tableau{4}}}{[40000]}}
&\oplusr \tabstack{\widetilde{\tableau{2}}}{[20000]} 
&\phantom{\oplusr \tabstack{\cdot}{[00000]}}
\\[1ex]
\lefttabbrack(\tabstackmed{\tableau{1 1 1 1 1 1}}{[00011]} \otimesr
\tabstackmed{\tableau{1 1 1 1 1 1}}{[00011]} \righttabbrack)_{\!\!\!s}  
&\raise2ex\hbox{$\rightarrow$}&
\tabstack{\widetilde{\tableau{2 2 2 2}}}{[00022]}
&\oplusr \tabstack{\widetilde{\tableau{2 2 2}}}{[00200]}
&\phantom{\oplusr \tabstack{\widetilde{\tableau{3 1 1 1}}}{[20011]}}
&\oplusr \tabstack{\tableau{1 1 1 1}}{[00011]} 
&\oplusr \tabstack{\widetilde{\tableau{2 2}}}{[02000]} 
&\phantom{\oplusr \tabstack{\widetilde{\tableau{4}}}{[40000]}}
&\oplusr \tabstack{\widetilde{\tableau{2}}}{[20000]} 
&\oplusr \tabstack{\cdot}{[00000]}\,
\end{align}
\end{subequations}
% Numbers below the reps indicate the rank of the corresponding tableau(x):
%
%     1X[0,0,0,0,4] +1X[0,0,0,2,2] +2X[0,0,2,0,0] +1X[0,1,0,1,1] +
%                         8             6 + 10
%
%     1X[0,1,1,0,2] +1X[0,2,0,0,0] +1X[1,0,0,0,2] +1X[1,0,0,1,3] +
%                         4
%
%     1X[1,0,0,2,0] +1X[1,0,1,1,1] +1X[1,1,0,0,2] +1X[1,1,1,0,0] +
%
%
%     1X[2,0,0,0,0] +1X[2,0,0,0,4] +1X[2,0,0,1,1] +1X[2,0,2,0,0] +
%           2                             6
%
%     1X[2,1,0,0,0] +1X[3,0,0,0,2] +1X[4,0,0,0,0]
%                                         4
%
%----
%
%     1X[0,0,0,0,4] +1X[0,0,0,1,1] +1X[0,0,0,2,2] +2X[0,0,1,0,2] +
%                          4               8
%  
%     1X[0,0,2,0,0] +1X[0,1,0,0,0] +2X[0,1,0,1,1] +1X[0,2,0,0,0] +
%           6                                            4
%  
%     2X[1,0,0,0,2] +1X[1,0,0,1,3] +2X[1,0,1,0,0] +1X[1,0,1,1,1] +
%  
%  
%     1X[1,1,0,0,2] +1X[1,1,1,0,0] +1X[2,0,0,0,0] +1X[2,0,0,1,1] +
%                                         2             6
%  
%     1X[2,1,0,0,0]

When deriving these decompositions, one has to pay attention to the
fact that representations sometimes can be realised in terms of two
different Young tableaux. This happens because of the presence of the
invariant epsilon tensor. The main example that appears in the
decomposition above is
\begin{equation}
\label{e:onerep_twotableaux}
\lefttabbrack(\tabstackhigh{\widetilde{\tableau{2 1 1 1 1}}^+}{[10002]} \otimesr
\tabstackhigh{\widetilde{\tableau{2 1 1 1 1}}^+}{[10002]}
\righttabbrack)_{\!\!\!s}  \raise2ex\hbox{$=$}
\raise2.5ex\hbox{\;\;\ldots\;\;} \oplusr \tabstackhigh{\widetilde{\tableau{2 2 2}}}{[00200]} 
\oplusr
\tabstackhigh{\widetilde{\tableau{2 2 2 1 1 1 1}}}{[00200]} 
\oplusr \raise2.5ex\hbox{\;\;\ldots\,.}
\end{equation}
Only the first tableau is compatible with the structure of the string 
amplitude, which explains why the $[00200]$ only occurs with
unit multiplicity on the first line of~\eqn{e:DF5squareexpansions}.

The invariant action should now be expressible in terms of a basis of scalars
constructed from tensor products of the $W^2$ and $(DF_{(5)}^+)^2$
representations.  Inspecting~\eqn{e:W2reps}
and~\eqn{e:DF5squareexpansions}, this gives a total of twenty-one
terms, of which seven remain when the Bianchi identities are imposed.
A list of these seven on-shell invariants, i.e.~the tensor
contractions corresponding to the Young tableaux, is constructed in
appendix~\ref{s:polynomials} (in our calculations we have, however,
kept track of all twenty-one invariants as a check on signs and
factors, and to have an efficient way to eliminate terms proportional
to Bianchi identities).  In order to write down the full on-shell
superinvariant with two five-form tensors that appears in string
theory, we then only have to give the numerical values of the \emph{seven}
coefficients of these building blocks. These coefficients are computed
in the following section, and the final result can be found in
equation~\eqn{e:finalresult}.
\end{sectionunit}

\begin{sectionunit}
\title{Amplitude results and the effective action}
\maketitle
\label{s:amplituderesults}

All ingredients that are necessary to extract an effective action from
the string amplitude~\eqn{e:fourpointampl} are now
available. Inserting the correlators computed in
section~\ref{s:stringstuff}, doing the (trivial) integral over the
modular parameter and performing the spinorial traces, we end up with
a result which is to be reproduced by a yet-to-be-determined effective
action. The latter can be expressed in the group theory basis
discussed in the previous section. 

In general, constructing an effective action from a set of string
amplitudes is a non-trivial procedure. One starts at the lowest order
in the external fields at which string theory produces a non-zero
amplitude, and constructs a corresponding term in the effective action
which reproduces this result. Iterating this procedure with more and
more external states in the string amplitude, one generically finds
that the corresponding field theory amplitude receives contributions
from all lower order terms in the action. This so-called field-theory
subtraction problem is, fortunately, absent for our four-point
calculation. This is because of the fact that there are no
non-vanishing three-point amplitudes at genus one. The entire
four-point amplitude must therefore be generated by a new four-field
term in the effective action. A related problem concerns the
contribution to the field theory amplitude of vertices in the
effective action which appear at lower order in the derivatives. Again
we find that, due to the low number of external states, this issue
does not cause any problems in the present situation.

All this means that we can transcribe the amplitude in a relatively
straightforward way to an effective action. Just before doing the
traces, this leads to
\begin{equation}
\label{e:theamplitude}
\begin{aligned}
\mathcal{L}  & = W_{r_1r_2s_1s_2} W_{r_3r_4s_3s_4} \\[1ex]
& \times \Tr\begin{aligned}[t]\bigg\{
 &D_{m} \slashed{F}^+
 \Big( \Gamma^{r_1\ldots r_4 m} 
      + 4\, \Gamma^{[r_1r_2r_3} \eta^{r_4] m}
      + 8\, \Gamma^{r_1} \eta^{r_2r_3}\eta^{r_4 m}
      + 8\, \Gamma^{r_3} \eta^{r_4r_1}\eta^{r_2 m}
      + 2\, \Gamma^{m} \eta^{r_1r_3} \eta^{r_2r_4}
 \Big)\\[1ex]
\times & D_{n} \slashed{F}^+
 \Big( \Gamma^{s_1\ldots s_4 n} 
      + 4\, \Gamma^{[s_1s_2s_3} \eta^{s_4] n}
      + 8\, \Gamma^{s_1} \eta^{s_2s_3}\eta^{s_4 n}
      + 8\, \Gamma^{s_3} \eta^{s_4s_1}\eta^{s_2 n}
      + 2\, \Gamma^{n} \eta^{s_1s_3} \eta^{s_2s_4}
 \Big)\bigg\}
\end{aligned}
\end{aligned}
\end{equation}
Here we have used the fact that the two $\Gamma^{[3]}$ terms on the
second and third line of Table~\ref{t:fourpointbosons} combine to form
a tensor fully antisymmetric in $r_1,\ldots,r_4$ (and similarly in
$s_1,\ldots,s_4$ for the right-moving sector). Observe that all terms
come with a plus sign, despite the appearance of the gamma-matrix
transpose in~\eqn{e:fourpointampl} for the anti-holomorphic
sector. Rather than explaining from first principles how these signs
arise, we will instead argue for their correctness in
section~\ref{s:threeforms} by observing that this is the only
combination which leads to a result consistent with the type-IIB
SL(2,$\mathbb{Z}$) duality symmetry. Alternatively, observe that these
signs are the only ones leading to an amplitude which is symmetric
under exchange of the external five-form states.

Before we proceed to rewrite this trace in a more usable form, we
should make a few comments. Firstly, one may wonder how self-duality
of the five-form polarisation tensors---a direct consequence of the
Weyl-spinor contraction of $\Gamma^{[5]}$ in the vertex
operator~\eqn{e:fiveform_vop}---is to be dealt with in the actual
calculations.  Our approach has been to work with manifestly self-dual
five-forms, i.e., to let the projection operator
$\mathcal{P}_+=\tfrac{1}{2}(\mathbb{1}+*)$ act on each five-form field
strength. For the amplitude trace in~\eqn{e:fourpointampl}
(or~\eqn{e:theamplitude}), it is straightforward to show that this has
the effect of producing an overall factor $\Tr(\Pi_+)$, where
$\Pi_\pm=\tfrac{1}{2}(\mathbb{1}\pm\Gamma^{\#})$ denotes the Weyl
projection operators. Since the trace is over Weyl spinors with a U(1)
R-charge, this results in an overall multiplicative factor of 32. In
addition, there is another overall multiplicative factor of 2,
resulting from adding the parity-even and -odd parts of the trace
(which become identical after imposing self-duality).

Secondly, we should comment on the dependence of the effective action
on Ricci tensor and scalar factors. When expanded to linear order in
the fluctuations these factors vanish by virtue of the on-shell
conditions on the vertex operator polarisation tensors. Therefore, one
would in principle have to analyse higher-point amplitudes to
determine the dependence on these tensors.  For applications in which
the background is taken to be Ricci-flat at lowest order in $\alpha'$,
the fact that these terms are not known may, however, not necessarily
pose a problem.\footnote{We adopt here the point of view that we
first determine the effective action in a fixed basis of the fields,
corresponding to the string theory vertex operators, and only
afterwards analyse how field redefinitions might simplify this
action. It is impossible to analyse the latter issue when the
dependence on Ricci tensor and scalar terms is unknown, as these are
related to $(F_{(5)}^+)^2$ type terms of rank two by the lowest-order
equations of motion.}

The spinorial traces that are left in~\eqn{e:theamplitude} are rather
complicated and lead to a plethora of terms without any obvious
structure. As announced, the most systematic way to organise the
result is to use the group-theory basis constructed in
section~\ref{s:invclass}. Expressed in this way we find the following
expression for the four-field effective action at order $(\alpha')^3$,
at up to two powers of the five-form field strength and up to terms
appearing in the lowest-order equations of motion:
% The traces are calculated in W3F2-rev5-trace55sym.nb,
% --trace33sym.nb and --trace11.nb.
\begin{multline}
\label{e:finalresult}
S^{W^2 (DF_{(5)}^+)^2}_{\text{IIB}} = \int\!{\rm d}^{10}x\,\sqrt{-g}\;
(DF_{(5)}^+\Big|_{\widetilde{\tableau{2 1 1 1 1}}^+})^2 \\[1ex]
\times \bigg( \begin{aligned}[t]
  &(16+\lambda)\, W^2\Big|_{\widetilde{\tableau{2 2 2 2}}} 
- 4(16-\lambda)\, W^2\Big|_{\widetilde{\tableau{2 2 2}}}
+192\, W^2\Big|_{\widetilde{\tableau{3 1 1 1}}}\\[1ex]
+&\tfrac{16}{15}(16+\lambda)\, W^2\Big|_{\widetilde{\tableau{2 2}}_A}
+\tfrac{32}{3}\,   W^2\Big|_{\widetilde{\tableau{4}}}
+\tfrac{1}{21}(16+\lambda)\, W^2\Big|_{\widetilde{\tableau{2}}}\bigg)\,,
\end{aligned}
\end{multline}
where we have ignored an overall normalisation constant.  The
coefficient of the remaining basis element $\widetilde{\tableau{2
2}}_S$ vanishes identically.  The explicit index contractions of the
various terms can, if necessary, be obtained from the expressions
given in appendix~\ref{s:polynomials}.

Moreover, we have here introduced a constant $\lambda$ to parameterise
an ambiguity in the four-field action as determined purely from a
four-point string-amplitude calculation. This ambiguity is directly
analogous to the one at four-point level in the coefficient of the
$\epsilon_{10}\epsilon_{10}W^4$ term: both the latter and the part of
the action proportional to $\lambda$ are total derivatives at
linearised level, and their respective coefficients can therefore not
be reliably determined by four-point calculations.\footnote{The
ambiguous part of the action was determined by linearising a general
linear combination of the seven basis invariants and subsequently
imposing on-shell identities and momentum conservation, and finally
requiring the resulting expression to vanish.}  While the coefficient
of the $\epsilon_{10}\epsilon_{10}W^4$ action was fixed
by~\dcite{gris11} by means of a four-loop sigma-model $\beta$-function
calculation, it could alternatively have been determined by the
calculation of suitable higher-point string amplitudes. We will not
attempt to determine $\lambda$ by calculating a five-point string
amplitude here. In the discussion, we shall however comment on a very
suggestive way to fix its value based on the linear scalar superfield.

We should stress that the fact that we have restricted the higher-derivative 
action to depend on only the self-dual part of the five-form field strength 
causes no problems. Rather, it is the correct thing to do, and leads, upon
varying the action~\eqn{e:finalresult} with respect to the four-form gauge 
potential $C_4$, to an~$\alpha'$-corrected self-duality relation for the
composite field strength
\begin{equation}
\tilde{F}_5 := {\rm d}C_4 +\tfrac{1}{2} B_2\wedge F_3 
  - \tfrac{1}{2} H_3 \wedge C_2
\end{equation}
(here $H_3={\rm d}B_2$ and $F_3={\rm d} C_2$). To see how this comes about,
first recall that the equation of motion for $C_4$ to lowest order, i.e.~the 
self-duality condition ${*}\tilde F_5=\tilde F_5$, cannot be obtained from a 
covariant action functional~\cite{Marcus:1982yu}. The best we can do is to 
write down an action $S^{(0)}$ which is \emph{compatible} with self-duality, 
in the sense that the associated Euler--Lagrange equation for $C_4$ combined 
with the Bianchi identity for $\tilde{F}_5$ requires the latter to be self-dual
up to an exact form. Explicitly, the two equations read, respectively,
\begin{equation}
{\rm d} {*}\tilde{F}_5 = H_3\wedge F_3 \quad\text{and}\quad
{\rm d} \tilde{F}_5 = H_3\wedge F_3\, .
\end{equation}
This logic repeats itself at higher order in~$\alpha'$. The field equation 
for $C_4$ derived from the action including $S^{(0)}$ and our four-field
result~\eqn{e:finalresult}---or, more generally, the full order-$(\alpha')^3$ 
action $S^{(3)}$, which is assumed to depend on $C_4$ only through 
$\tilde F_5^+$ and $D\tilde F_5^+$---takes the form
\begin{equation}
\label{e:orderthree}
{\rm d} \left[{*}\tilde{F}_5 - 2\, \mathcal{P}_-\left(
        \frac{\delta S^{(3)}}{\delta\tilde F_5^+}\right)\right]
      = H_3\wedge F_3\, ,
\end{equation}
where $\mathcal{P}_-=\tfrac{1}{2}(1-*)$ is the projection operator onto
antiself-dual five-forms.
%where the five-form $X_5$ is defined by the relation 
%$\delta_{C_4} S^{(\alpha')^3} = \int X_5 \wedge \delta_{C_4} \tilde F_5^+$.
Combined with the Bianchi identity for $\tilde{F}_{(5)}$, this equation is 
compatible (in the sense discussed above) with the $\alpha'$-corrected 
self-duality relation
\begin{equation}
 \tilde{F}_5 + \frac{\delta S^{(3)}}{\delta\tilde F_5^+} =
 {*}\left[ \tilde{F}_5 + \frac{\delta S^{(3)}}{\delta\tilde F_5^+} 
   \right] \,.
\end{equation}
The presence of $\mathcal{P}_-$ in~\eqn{e:orderthree}, which is crucial for 
consistency, is a direct consequence of the fact that $\tilde F_5$ is projected
onto its self-dual part in $S^{(3)}$. 
(In the vertex operators that appear in our four-point calculations one 
only needs the linearised version of this constraint, which is of course 
simply stating that the polarisation tensor is self-dual.)

\end{sectionunit}

\begin{sectionunit}
\title{A cross-check using SL(2,$\mathbb{Z}$) invariance}
\maketitle
\label{s:threeforms}
The calculation for the five-form terms which have discussed so far
shares many aspects with the calculation one has to do in order to
determine the dependence of the effective action on the Ramond-Ramond
three-form field strength. The resulting~$R^2 (DF_{(3)})^2$ terms in
the effective action should, as we explain below, be related by
SL(2,$\mathbb{Z}$) symmetry to the~$R^2 (DH_{(3)})^2$ terms involving
the Neveu-Schwarz three-form field strength, which have been computed
by~\dcite{Gross:1987mw}.\footnote{At the level of the effective
supergravity theories we are discussing the symmetry group is
SL(2,$\mathbb{R}$), but we will refer to the smaller, discrete subgroup
that is the S-duality symmetry of the full type-IIB string theory.}
Matching their calculation with ours leads to an interesting check of
the duality symmetry of the type-IIB theory, while at the same time
providing a very strong check on all signs and factors in our string
calculation.

To explain the logic, recall that~SL(2,$\mathbb{Z}$) symmetry dictates
that the three-form field strengths should appear in the effective
action only through the SL(2,$\mathbb{Z}$)-invariant combination
\begin{equation}
\label{e:Gdef}
G_{(3)} = \frac{1}{\sqrt{\Imag{\tau}}}\big( F_{(3)} + \tau H_{(3)}\big)\, ,
\end{equation}
with U(1) charge one, as well as its complex conjugate. The complex
coupling~$\tau$ is formed from the Ramond-Ramond scalar and the
dilaton,
\begin{equation}
\tau = \chi + i e^{-\phi}\, .
\end{equation}
Terms in the effective action are then built from~\eqn{e:Gdef} as well
as other SL(2,$\mathbb{Z}$) singlets, whose transformation behaviour
under the local~U(1) symmetry is compensated for by pre-factor modular
functions~$f^{(w,-w)}$ of the complex coupling~$\tau$ (see
e.g.~\dcite{Green:1998by} for more details).

The terms computed by~\dcite{Gross:1987mw} are those independent of
$\chi$, i.e.~for which \mbox{$\tau=i\Imag\tau$}. We will also restrict
ourselves to this case, because we did not compute string amplitudes
with an external Ramond-Ramond scalar. In the Einstein frame, the NS-NS 
field-strength part of the genus-zero 
effective action is then of the
form\footnote{We remind the reader that under a super-Weyl rescaling
from the Einstein frame to the string frame the metric transforms as
$g^E_{\mu\nu} = e^{-\phi/2} g^S_{\mu\nu}$ and that this induces the
transformation
\begin{multline}
\int\!\rmd^{10}x \, \sqrt{- g}\, \Big[  R +  
(\alpha')^3 e^{-3\phi/2}  t_8 t_8  R^4
- \tfrac{1}{2}(\partial_\mu\phi)^2
\Big]\bigg|_{\text{Einstein}} \\[1ex]
= \int\!\rmd^{10}x\, \sqrt{-g} \, e^{-2\phi} \Big[ R +
(\alpha')^3 t_8 t_8 R^4
+ 4(\partial_\mu\phi)^2 \Big]\bigg|_{\text{string}}
\end{multline}
on the genus-zero string effective action.}
\begin{equation}
e^{-5\phi/2} \sqrt{-g} R^2 (DH_{(3)})^2 \Big|_{\text{Einstein}}
\rightarrow 
e^{-2\phi} \sqrt{-g} R^2 (DH_{(3)})^2 \Big|_{\text{string}}\, .
\end{equation}
with the specific tensorial structure given by~\dcite{Gross:1987mw}.
Their result can be made SL(2,$\mathbb{Z}$) invariant in two ways,
\begin{equation}
\begin{aligned}
 e^{-3\phi/2} R^2 DG_{(3)} DG^*_{(3)}\Big|_{\text{Einstein}}
&= e^{-5\phi/2} R^2 (DH_{(3)})^2 + e^{-\phi/2} R^2 (DF_{(3)})^2\Big|_{\text{Einstein}}\, ,\\[1ex]
-e^{-3\phi/2} R^2\, \tfrac{1}{2}\big( DG_{(3)} DG_{(3)} + \text{c.c.}\big) \Big|_{\text{Einstein}}
&= e^{-5\phi/2} R^2 (DH_{(3)})^2 - e^{-\phi/2} R^2 (DF_{(3)})^2\Big|_{\text{Einstein}}\, .
\end{aligned}
\end{equation}
We thus find that the string effective action should contain terms of the
form
\begin{equation}
\label{e:GS2predict}
R^2 (DF_{(3)})^2\Big|_{\text{string}}\, ,
\end{equation}
with tensorial structures identical to those of the Neveu-Schwarz
dependent terms. Moreover, by the general logic sketched above, terms
with this structure appear both at tree level and at genus one.
The relative coefficient with respect to the Neveu-Schwarz sector
depends on the unknown relative normalisation between the two type of
terms given above (this coefficient has been conjectured
by~\dcite{Kehagias:1998cq} but an explicit derivation of it is
lacking).

The most basic way in which the above prediction of the duality symmetry 
can be verified is by comparing the amplitude with two gravitons and two 
external NS-NS three-form states to the one with two gravitons and two 
external RR states. The most interesting aspect of this comparison is
that the fermionic world-sheet correlators that appear in these two
calculations are completely different. In the NS-NS sector (for full
details we refer to~\dcite{Gross:1987mw}) the relevant correlator is
formed from RNS fermions only, taking the form
\begin{equation}
\langle
\norder{\Psi\Psi}{(z_1)}\norder{\Psi\Psi}{(z_2)}
\norder{\Psi\Psi}{(z_3)}\norder{\Psi\Psi}{(z_4)}
\rangle
\end{equation}
in each of the chiral sectors.  The correlator in the RR sector, on
the other hand, contains spin fields which lead to traces over gamma
matrices, as we have explained in detail in section~\ref{s:stringstuff}. 
We have compared the long expressions for the $ggH_{(3)}H_{(3)}$ and 
$gg F_{(3)}F_{(3)}$ amplitudes at eighth order in the momenta.  
Remarkably, after taking into account the on-shell conditions on the 
polarisation tensors and imposing momentum conservation, the two 
amplitudes (each with several hundred different terms) agree perfectly!

In order to establish the SL(2,$\mathbb{Z}$) duality at the level of the 
full covariant effective action, one would need to have precise control 
over the terms in the action that lead to a vanishing on-shell four-point 
function. Since the five-point $gggH_{(3)}H_{(3)}$ amplitude has never 
been fully computed in the literature, these terms are ambiguous already 
in the NS-NS sector.\footnote{There is a widely expressed belief that the 
$W^2DH_{(3)}^2$ terms can simply be obtained from the $W^4$ terms by using 
a generalised spin connection modified by the addition of a torsion term 
in the form of the NS-NS three-form field strength. There is, however, no 
proof that this mechanism works in general (see, e.g., \dcite{Metsaev:1987zx}
for a discussion) and we will therefore not rely on it here.} In order to 
enable a systematic future comparison, we have therefore again decomposed 
the effective action in a basis of irreducible invariants and parameterised 
the ambiguity by determining those linear combinations of the basis 
invariants that lead to vanishing four-point functions.

This decomposition is similar in spirit to the one given in the
previous section for the five-form terms. The three-form derivatives
come in three different irreducible representations:
\begin{equation}
\raise2ex\hbox{$DF_{(3)}:$} \quad \tabstack{\tableau{1}}{[10000]} \otimesr
\tabstack{\tableau{1 1 1}}{[00100]}  \raise2ex\hbox{$=$}
\tabstack{\widetilde{\tableau{2 1 1}}}{[10100]}
\oplusr \tabstack{\tableau{1 1 1 1}}{[00011]}
\oplusr \tabstack{\tableau{1 1}}{[01000]}\, \raise2ex\hbox{.}
\end{equation}
We can immediately discard the $[01000]$ as it corresponds to the 
lowest-order equation-of-motion term $D^m F_{mnp}$, which we simply set 
to zero throughout the calculations. The products of the remaining two 
$DF_{(3)}$ (or $DH_{(3)}$) representations have the decompositions
\begin{subequations}
\label{e:DF3squareexpansions}
\begin{align}
\left(\tabstack{\widetilde{\tableau{2 1 1}}}{[10100]} \otimesr
\tabstack{\widetilde{\tableau{2 1 1}}}{[10100]} \right)_{\!\!\!s}  \raise2ex\hbox{$\rightarrow$}&
\tabstack{\widetilde{\tableau{2 2 2 2}}}{[00022]}
&\oplusr \tabstack{2\,\widetilde{\tableau{2 2 2}}}{[00200]}
&\oplusr \tabstack{2\,\widetilde{\tableau{3 1 1 1}}}{[20011]}
&\oplusr \tabstack{2\,\tableau{1 1 1 1}}{[00011]} 
&\oplusr \tabstack{3\,\widetilde{\tableau{2 2}}}{[02000]} 
&\oplusr \tabstack{\widetilde{\tableau{4}}}{[40000]} 
&\oplusr \tabstack{2\,\widetilde{\tableau{2}}}{[20000]} 
\oplusr \tabstack{\cdot}{[00000]}\,
\end{align}
\begin{align}
%\\[1ex]
\left(\tabstack{\widetilde{\tableau{2 1 1}}}{[10100]} \otimesr
\tabstack{\tableau{1 1 1 1}}{[00011]} \right)  \raise2ex\hbox{$\rightarrow$}&
&\phantom{\oplusr} \tabstack{\widetilde{\tableau{2 2 2}}}{[00200]}
&\oplusr \tabstack{\widetilde{\tableau{3 1 1 1}}}{[20011]}
&\oplusr \tabstack{\tableau{1 1 1 1}}{[00011]} 
&\oplusr \tabstack{\widetilde{\tableau{2 2}}}{[02000]} 
&
&\oplusr \tabstack{\widetilde{\tableau{2}}}{[20000]} 
\\[1ex]
\left(\tabstack{\tableau{1 1 1 1}}{[00011]} \otimesr
\tabstack{\tableau{1 1 1 1}}{[00011]} \right)_{\!\!\!s}  \raise2ex\hbox{$\rightarrow$}&
\tabstack{\widetilde{\tableau{2 2 2 2}}}{[00022]}
&\oplusr \tabstack{\widetilde{\tableau{2 2 2}}}{[00200]}
&
&\oplusr \tabstack{\tableau{1 1 1 1}}{[00011]} 
&\oplusr \tabstack{\widetilde{\tableau{2 2}}}{[02000]} 
&
&\oplusr \tabstack{\widetilde{\tableau{2}}}{[20000]} 
 \oplusr \tabstack{\cdot}{[00000]}\,
\end{align}
\end{subequations}
Here we have again left out representations not present in the
expansion of two Weyl tensors (given in~\eqn{e:W2reps}) as well as
terms which are not compatible with the symmetries of the tensorial
structures listed in table~\ref{t:twotwo}.\footnote{Once more, one has
to be careful to recognise that one representation can appear as two
different Young tableaux; see the discussion
around~\eqn{e:onerep_twotableaux}. In the present case, this happens
for~the $[20011]$ representation in the second line
of~\eqn{e:DF3squareexpansions}.}

Verifying SL(2,$\mathbb{Z}$) symmetry at the level of the effective
actions amounts to establishing that the respective expansions in basis
invariants for $W^2(DH_{(3)})^2$ and $W^2(DF_{(3)})^2$ agree when all 
terms proportional to $DH_{(3)}$ or $DF_{(3)}$ in either the $[01000]$ 
or the $[00011]$ representation have been set to zero.  Combining the 
$W^2$ expansion~\eqn{e:W2reps} with the first line 
of~\eqn{e:DF3squareexpansions}, one finds that this corresponds to a 
matching of 17 coefficients.

The decomposition of the effective action on the group-theory basis is
thus more complicated than in the case of the five-form action. The three
invariants based on the $\widetilde{\tableau{2 2}}_S$ representation again 
do not appear, while the terms that lead to a vanishing four-point amplitude
form a four-parameter family of ambiguous terms parameterised by
$\lambda_1\ldots\lambda_4$. Including these ambiguous terms, the expression 
for the action at four-point level reads
\begin{multline}
\label{e:finalresult3}
S^{W^2 DF_{(3)}^2}_{\text{IIB}} = 
 \int\!{\rm d}^{10}x\,\sqrt{-g}\;
 (DF_{(3)}\Big|_{\widetilde{\tableau{2 1 1}}})^2 \\[1ex]
\times \begin{aligned}[t]
    \bigg( 
&  (-240+\lambda_1)                           \, W^2\Big|_{\widetilde{\tableau{2 2 2 2}}} 
 + (-64 +\lambda_2)                           \, W^2\Big|_{\widetilde{\tableau{2 2 2}}_{a}}\\[1ex]
&+ 12\,(32 +\tfrac{2}{5}\lambda_1 - \lambda_2)\, W^2\Big|_{\widetilde{\tableau{2 2 2}}_{b}}
 - 96 \, W^2\Big|_{\widetilde{\tableau{3 1 1 1}}_{a}}
 + 96 \, W^2\Big|_{\widetilde{\tableau{3 1 1 1}}_{b}} \\[1ex] 
&- 64 \, W^2\Big|_{\widetilde{\tableau{4}}} 
 + (32+\lambda_4)                                    \, W^2\Big|_{\tableau{1 1 1 1}_{a}} 
 - 32 \, W^2\Big|_{\tableau{1 1 1 1}_{b}} 
 + (\tfrac{96}{5}+\lambda_3)                  \, W^2\Big|_{\widetilde{\tableau{2 2}}_{Aa}} \\[1ex]
&+ (\tfrac{256}{5}+\tfrac{32}{25}\lambda_1 - 2\lambda_2 -\tfrac{4}{3}\lambda_3)
                                                \, W^2\Big|_{\widetilde{\tableau{2 2}}_{Ab}} 
 + \tfrac{8}{5}\,(-32-\tfrac{1}{5}\lambda_1)   \, W^2\Big|_{\widetilde{\tableau{2 2}}_{Ac}} \\[1ex]
&+ (-\tfrac{16}{105}\lambda_1+\tfrac{25}{84}\lambda_2+\tfrac{5}{18}\lambda_3)
                                                \, W^2\Big|_{\widetilde{\tableau{2}}_a} 
 + (\tfrac{8}{7}+\tfrac{1}{35}\lambda_1-\tfrac{5}{84}\lambda_2-\tfrac{5}{36}\lambda_3)
                                                \, W^2\Big|_{\widetilde{\tableau{2}}_b} \\[1ex]
&+ (\tfrac{68}{315}-\tfrac{11}{3150}\lambda_1+\tfrac{1}{144}\lambda_2+\tfrac{1}{108}\lambda_3)
                                                \,W^2\Big|_{\cdot}
   \bigg)\, .
\end{aligned}
\end{multline}
We have again discarded an overall normalisation constant.  The indices on 
the tableaux serve to distinguish inequivalent ways of constructing terms 
with the same Young symmetries.  We refrain from giving the explicit 
expressions for these 14 basis invariants here, as the SL(2,$\mathbb{Z}$) 
symmetry in this case provides us with a more compact way of writing the 
action: by a suitable choice of the parameters $\lambda_i$, the action 
given above can be reduced to the form \begin{equation}
\label{e:t8t8_min_epseps}
(t_8t_8+\tfrac{1}{8}\epsilon_{10}\epsilon_{10})\, W^2 (DF_{(3)})^2\,.
\end{equation}
The parameter values corresponding to this action are given by 
$\lambda_1=240$, $\lambda_2=192$, $\lambda_3=-576/5$ and $\lambda_4=0$.
In particular, they have been adjusted so that the rank-eight contractions 
between the $W^2$ and $DF_{(3)}^2$ blocks cancel between the two terms 
in~\eqn{e:t8t8_min_epseps}. This gives precisely the RR three-form 
version of the action obtained by inserting a shifted spin connection 
\mbox{$\tilde \omega = \omega + H_{(3)}$} in the familiar $W^4$ action of the 
type-IIB theory, showing that our result~\eqn{e:finalresult3} is
compatible also with a stronger check of the SL(2,$\mathbb{Z}$) 
duality symmetry.\footnote{As a curiosity, let us mention that the 
above values of the $\lambda$ parameters correspond to the 
$\epsilon_{10}\epsilon_{10}$ part of~\eqn{e:t8t8_min_epseps} multiplied
by a factor of two. Hence, the action~\eqn{e:finalresult3} with all
$\lambda_i$ set to zero also takes the form~\eqn{e:t8t8_min_epseps} but
with the opposite sign for the second term.}

\end{sectionunit}
\end{sectionunit}

\begin{sectionunit}
\title{Discussion and comments on applications} 
\maketitle

\begin{sectionunit}
\title{Comparison with predictions of the linearised superfield}
\maketitle

Berkovits and Howe have recently shown that the linearised scalar
superfield of the type-IIB theory~\cite{nils1,howe1} does not admit a
non-linear extension, essentially because no chiral measure can be
constructed (a detailed argument can be found
in~\dcite{deHaro:2002vk}). In the present section we would like to
make a few concluding remarks relating our four-point amplitudes to
the predictions of the \emph{linearised} superfield. We would,
however, like to stress that, despite the comments we will make in
this section, the direct string amplitude computations of the type
presented in our paper are currently the only known way of
constructing the full dependence of the type-IIB invariant on the
Ramond-Ramond gauge fields.

In the linearised superfield approach, the action arises as a
sixteen-dimensional fermionic integral over four powers of the scalar
superfield $\Phi$,
\begin{equation}
\label{e:IIBintegral}
S_{\text{IIB}} = \int\!{\rm d}^{10}x\,{\rm d}^{16}\theta\, e\,
\Phi^4\,.
\end{equation}
This integral is very hard to do explicitly except
when one restricts to pure graviton terms, but we can again, as
in~\cite{Peeters:2001ub}, use arguments based on representation theory
to compare with our string-based result. The linearised scalar
superfield contains, at fourth order in~$\theta$, the terms
\begin{equation}
\Phi = \cdots + (\theta{\cal C} \Gamma^{r_1r_2r_3} \theta)\,
(\theta{\cal C} \Gamma^{s_1 s_2 s_3}\theta)\,
\Big( R_{r_1 r_2 s_1 s_2} \eta_{r_3 s_3} + D_{r_1} F_{r_2 r_3 s_1 s_2 s_3} 
   \Big) + \cdots \, .
\end{equation}
The four powers of the anti-commuting~$\theta$ restrict the
fields to the representations
\begin{equation}
\label{e:sufielddecomp}
\raise2ex\hbox{$(\otimes_{i=1}^4 \mathbf{16} )_a =$} 
  \tabstack{\widetilde{\tableau{2 2}}}{[02000]} \oplusr
  \tabstack{\widetilde{\tableau{2 1 1 1 1}}^+}{[10002]} \, \raise2ex\hbox{,}
\end{equation}
where the tensors on the right-hand side correspond to the Weyl tensor
and the non-trivial representation of $D F_{(5)}^+$, respectively. A
similar argument restricts the way in which~$W^2$ and $(DF_{(5)}^+)^2$
tensors can appear in the action. Namely, at level $\theta^8$ in
$\Phi^2$ one finds the decomposition
\begin{equation}
\label{e:sufieldsquaredecomp}
\raise2ex\hbox{$( \otimes_{i=1}^8 \mathbf{16} )_a =$} 
\tabstack{\widetilde{\tableau{2 2 2}}}{[00200]} \oplusr 
\tabstack{\widetilde{\tableau{3 1 1 1}}}{[20011]} \oplusr
\tabstack{\widetilde{\tableau{4}}}{[40000]} \, \raise2ex\hbox{.}
\end{equation}
We observe that this is, in general, more restrictive than the action 
given in~\eqn{e:finalresult}. 

However, as we have noted in section~\ref{s:amplituderesults},
there is a one-parameter ambiguity in our effective
action~\eqn{e:finalresult} which arises because the four-point amplitude 
does not suffice to fix the coefficients of terms in the action that lead 
to vanishing on-shell four-point functions. This ambiguity,
parameterised by $\lambda$ in~\eqn{e:finalresult}, is precisely such
that the action can be reduced to the three representations given
in~\eqn{e:sufieldsquaredecomp} above: for the special value
$\lambda=-16$ only these representations survive.  Assuming that the
superfield integral~\eqn{e:IIBintegral} produces the string
superinvariant---and the fact that there exists a value for $\lambda$
for which the action satisfies the restriction imposed 
by~\eqn{e:sufieldsquaredecomp} indeed lends a certain credence to this 
assumption---we can thus use this argument to fix the four-field 
$R^2 (DF_{(5)}^+)^2$ action completely. 

As a side note, the above decomposition also explains the precise
relative coefficient between the $t_8t_8 W^4$ and $\epsilon_{10}\epsilon_{10}
W^4$ parts of the superfield $W^4$ invariant (which happens to agree
with the corresponding part of the type-IIB four-point effective
action): there is a unique linear combination of the two that,
like~\eqn{e:sufieldsquaredecomp}, contains no rank-eight contractions
between two~$W^2$ factors.\footnote{Here one should again take note of
the discussion around~\eqn{e:onerep_twotableaux}. For the case at
hand, the irreducible representation $[20011]$ has a dual realisation
in terms of a rank-eight tensor. However, this tensor is
antisymmetric in six indices and can therefore not be be formed out
of two Weyl tensors.}
\end{sectionunit}

\begin{sectionunit}
\title{Applications, conclusions and outlook} \maketitle

We have shown how systematic string-amplitude calculations and a
group-theoretical approach to the construction of field-theory
invariants can be used to successfully compute the string effective
action including Ramond-Ramond fields. This is a technically rather
demanding project, but we have shown that the results can be cast in a
very compact form. We have illustrated our methods by giving an
explicit expression for the four-field terms in the type-IIB effective
action involving two powers of the three-form or five-form field
strength. Clearly, our work is a necessary prerequisite for any future
calculations of this type that extend to higher order in the
fields. For this reason, we consider it very important that our
calculations passed the rather spectacular SL(2,$\mathbb{Z}$) match.

One particularly interesting term in the effective action that can be
determined using our setup is the $W^3 (F_{(5)}^+)^2$ term; in fact,
the results obtained in the present paper are crucial in order to
determine these terms. We have done some preliminary work to
investigate the complexity of this problem. Firstly, while the string
amplitudes now contain considerably more terms, the correlators are
again all ``topological'' in the sense that they reduce to constants
by virtue of the Riemann identity. Secondly, one now encounters
field-theory subtraction issues: the five-point field theory amplitude
receives contributions from tree-level graphs formed from one
lowest-order supergravity vertex and one four-field vertex coming from
the $W^2 (DF_{(5)}^+)^2$ terms computed in the present paper. Despite
these complications, the effective action still has a manageable
expansion in group theory invariants. This is determined by the two
tensor products
\begin{subequations}
\begin{align}
\label{e:F2expansion}
\raise2ex\hbox{$\Big({\tableau{1 1 1 1 1}}^+\otimes\;{\tableau{1 1 1 1 1}}^+\Big)_s =$}&
\tabstack{\widetilde{\tableau{2}}}{[20000]} \;    \oplusr
\tabstack{\widetilde{\tableau{2 2 2}}}{[00200]} \oplusr 
\tabstack{\widetilde{\tableau{2 1 1 1 1}}^+}{[10002]} \oplusr
\tabstack{\widetilde{\tableau{2 2 2 2 2}}^+}{[00004]}\\[1ex]
\label{e:W3expansion}
\raise2ex\hbox{$\Big(\tilde{\tableau{2 2}}\otimes\tilde{\tableau{2 2}}
  \otimes\tilde{\tableau{2 2}} \Big)_s =$}&
  \tabstack{3\,\widetilde{\tableau{2}}}{[20000]} \; 
  \oplusr \;
  \tabstack{6\,\widetilde{\tableau{2 2 2}}}{[00200]} 
  \oplusr \;
  \tabstack{2\,\widetilde{\tableau{2 1 1 1 1}}^+}{[10002]} 
  \oplusr \;
  \tabstack{\cdots}{\phantom{3}}
\end{align}
\end{subequations}
where the ellipses denote representations that do not appear in the
expansion~\eqn{e:F2expansion}.  From the overlap between the two
expansions we see that there are, at most, eleven independent
$W^3\,(F_{(5)}^+)^2$ invariants. Computing this part of the type-IIB
action will be an interesting application of the procedure given in
the present paper.

Another set of terms which is related to the ones computed here are
those involving space-time fermions.  Part of our motivation for 
the paper~\cite{Peeters:2000qj} was to deduce the modifications to the
supersymmetry transformation rules and the resulting superalgebra in
order to determine the superspace torsion constraints. For this
purpose it is presumably sufficient to determine the fermion bilinear
terms in the effective action. 
\bigskip

Finally, let us comment on a more conceptual lesson which can be
learned from our analysis. Despite the very low number of basis
invariants from which our effective action is constructed, the
corresponding expressions as given in the appendix are quite lengthy
and the resulting action, when written out in this brute-force way,
would be rather intractable. It thus seems clear that any actual
applications along the lines of~\dcite{Frolov:2001xr}
or~\dcite{deHaro:2003zd} would benefit from keeping the Young
symmetrisers implicit as long as possible. 

A similar systematic approach is also highly desirable for the
determination of corrections to supersymmetry transformation rules. As
we have already shown in~\cite{Peeters:2000qj}, supersymmetry mixes
the tensorial structures of the various terms in the effective action
in a very complicated way. By writing not just the action but also the
supersymmetry transformation rules using a group-theory basis, it is
likely that one can cast the $(\alpha')^3$ corrections to these rules
in a manageable form, despite the underlying complexity of the
calculations. We will leave these issues for future work.

\end{sectionunit}

\end{sectionunit}

\section*{Acknowledgements}

We thank Stefan Theisen and Pierre Vanhove for discussions, and Paul
Howe for valuable remarks regarding the geometry of type-IIB
superspace.  For computation of the gamma-matrix traces, the
Mathematica package {\tt GAMMA} by~\dcite{Gran:2001yh} has been
helpful. Similarly, the programme {\tt LiE}~\cite{e_cohe1} has been
most useful for calculating SO(1,9) expansions.

\vfill\eject
\appendix
\begin{sectionunit}
\title{Appendix}
\maketitle
\begin{sectionunit}
\title{Explicit expressions for tensor polynomials}
\maketitle
\label{s:polynomials}
In this section we list the explicit expressions for the tensor
polynomials that appear in the effective action at order
$(\alpha')^3$. 

The first step in their construction is to decompose the $W^2$ and
$(DF_{(5)}^+)^2$ products in irreducible representations, as given
in~\eqn{e:W2reps} and~\eqn{e:DF5squareexpansions} of the main text.
The explicit forms of the Weyl tensor building blocks are given by
% Computed in:
%   W3F2-rev5-2x4window.nb
%   W3F2-rev5b-2x4window.nb
%   reps.cdb
\begin{align}
\label{e:w21}
W^2\Big|_{\widetilde{\tableau{2 2 2 2}}} &= 
\begin{aligned}[t]
&   W_{r_1 r_2 s_1 s_2} W_{r_3r_4 s_3s_4}
 -2\,\delta^{r_1}_{s_1} 
                  W_{d_1 r_2 s_2 s_3} W_{r_3 r_4 d_1 s_4}
 -\tfrac{4}{5} \delta^{r_1 r_2}_{s_1 s_2} 
                  W_{d_1 r_3 d_2 s_3} W_{d_1 s_4 d_2 r_4}\\[1ex]
 & +\tfrac{1}{5}  \delta^{r_1 r_2}_{s_1 s_2}
                  W_{d_1 d_2 r_3 r_4} W_{d_1 d_2 s_3 s_4}
 -\tfrac{2}{15}  \delta^{r_1 r_2 r_3}_{s_1 s_2 s_3}
                  W_{d_1 d_2 d_3 r_4} W_{d_1 d_2 d_3 s_4}\\[2ex]
 &+\tfrac{1}{210} \delta^{r_1 r_2 r_3 r_4}_{s_1 s_2 s_3 s_4}
                  W^2\, ,
\end{aligned}\\[2ex]
W^2\Big|_{\widetilde{\tableau{2 2 2}}} &= 
\begin{aligned}[t]
&  \tfrac{1}{2} W_{d_1 r_1 s_1 s_2} W_{d_1 s_3 r_2 r_3}
 +\tfrac{1}{3} \delta^{r_1}_{s_1} W_{d_1 r_2 d_2 s_2} W_{d_1 s_3 d_2 r_3}\\[1ex]
& +\tfrac{1}{14}\delta^{r_1 r_2}_{s_1 s_2} W_{d_1 d_2 d_3 r_3} W_{d_1
  d_2 d_3 s_3}
 -\tfrac{1}{12}\delta^{r_1}_{s_1} W_{d_1 d_2 r_2 r_3} W_{d_1 d_2 s_2
  s_3}\\[1ex]
& -\tfrac{1}{336}\delta^{r_1 r_2 r_3}_{s_1 s_2 s_3} W^2\, ,
\end{aligned}\\[2ex]
W^2\Big|_{\widetilde{\tableau{3 1 1 1}}} &= 
\begin{aligned}[t]
& W_{d_1 s_1 [r_1 r_2} W_{|d_1 s_2| r_3 r_4]} 
 -\tfrac{1}{12} \delta^{s_1s_2} W_{d_1 d_2 [r_1 r_2} W_{|d_1 d_2| r_3 r_4]}\\[1ex]
& +\tfrac{1}{18} \delta^{(s_1}{}_{[r_1} W_{|d_1 d_2| r_2 r_3} W_{|d_1 d_2| r_4]}{}^{s_2)}\\[1ex]
& +\tfrac{1}{9} \delta^{(s_1}{}_{[r_1} W_{|d_1 d_2| r_2}{}^{s_2)} W_{ d_1 d_2 | r_3 r_4]}\,,
\end{aligned}\\[2ex]
W^2\Big|_{\widetilde{\tableau{2 2}}_A} &= 
\begin{aligned}[t]
& \tfrac{2}{3} W_{d_1 r_1 d_2 r_2} W_{d_1 s_1 d_2 s_2}
 -\tfrac{2}{3} W_{d_1 r_1 d_2 s_1} W_{d_1 s_2 d_2 r_2}\\[1ex]
& -\tfrac{1}{4}\delta^{r_1}_{s_1} W_{d_1 d_2 d_3 r_2} W_{d_1 d_2 d_3 s_2}
  +\tfrac{1}{72} \delta^{r_1 r_2}_{s_1 s_2} W^2\, ,
\end{aligned}\\[2ex]
\label{e:w24}
W^2\Big|_{\widetilde{\tableau{2 2}}_S} &= 
\begin{aligned}[t]
& \tfrac{2}{3} W_{d_1 r_1 d_2 s_1} W_{d_1 r_2 d_2 s_2}
 +\tfrac{2}{3} W_{d_1 r_1 d_2 s_1} W_{d_1 s_2 d_2 r_2}\\[1ex]
& +\tfrac{1}{4}\delta^{r_1}_{s_1}W^{d_1 d_2 d_3 r_2}W_{d_1 d_2 d_3 s_2}
  -\tfrac{1}{72} \delta^{r_1 r_2}_{s_1 s_2} W^2\, ,
\end{aligned}\\[2ex]
W^2\Big|_{\widetilde{\tableau{4}}} & 
  = \, W_{d_1 (r_1| d_2 | r_2 } W_{| d_1 | s_1 | d_2 | s_2)} 
     - \tfrac{3}{14} \eta_{(r_1r_2} W_{s_1 | d_1 d_2 d_3 |}W_{s_2) d_1 d_2 d_3}
     + \tfrac{1}{112} \eta_{(r_1r_2} \eta_{s_1s_2)} W^2 \, , \\[2ex]
W^2\Big|_{\tableau{1 1 1 1}} & 
  = \, W_{d_1 [r_1 | d_2 | r_2 } W_{| d_1 | s_1 | d_2 | s_2]} \, , \\[2ex]
\label{e:w28}
W^2\Big|_{\widetilde{\tableau{2}}} & 
  = \, W_{d_1 d_2 d_3 r_1} W_{d_1 d_2 d_3 s_1} 
    - \tfrac{1}{10}\eta_{r_1s_1} W^2 \, .
\end{align}
These expressions are obtained by applying the Young-tableau
symmetrisers and subsequently subtracting traces in order to reduce to
an irreducible representation. Antisymmetry is assumed on the index
sets $r_i$ and $s_i$ except where otherwise explicitly indicated.
The non-trivial representation contained in $DF_{(5)}^+$ is given by
\begin{align}
\label{e:DF5explicit}
DF_{(5)}^+\Big|_{\widetilde{\tableau{2 1 1 1 1}}^+} &=
\begin{aligned}[t]
&  \frac{5}{6}  D_{r} F_{r_1 r_2 r_3 r_4 r_5}^+
 +\frac{1}{6} \Big( \begin{aligned}[t]
 &  D_{r_1} F_{r r_2 r_3 r_4 r_5}^+
 - D_{r_2} F_{r r_1 r_3 r_4 r_5}^+\\[1ex]
 & + D_{r_3} F_{r r_1 r_2 r_4 r_5}^+
 - D_{r_4} F_{r r_1 r_2 r_3 r_5}^+
 + D_{r_5} F_{r r_1 r_2 r_3 r_4}^+\Big)\end{aligned}\\[1ex]
&\quad - \frac{1}{10}\Big(
\begin{aligned}[t]
& \eta_{r r_1} D_{d_1} F_{d_1 r_2 r_3 r_4 r_5}^+
 -\eta_{r r_2} D_{d_1} F_{d_1 r_1 r_3 r_4 r_5}^+\\[1ex]
& -\eta_{r r_3} D_{d_1} F_{d_1 r_1 r_2 r_4 r_5}^+
 -\eta_{r r_4} D_{d_1} F_{d_1 r_1 r_2 r_3 r_5}^+
 -\eta_{r r_5} D_{d_1} F_{d_1 r_1 r_2 r_3 r_4}^+\Big) \, .\end{aligned}
\end{aligned}
\end{align}
There are \emph{a priori} two ways to contract two of these $DF_{(5)}^+$
factors in such a way as to obtain a tensor of rank eight. However, the
classification in section~\ref{s:invclass} has shown that there can be 
only one independent object of this type. Indeed, explicit calculation 
shows that one of these contractions vanishes:
% 
% W3F2-rev5b-twogamAr.nb for the zero and two-\epsilon terms,
% W3F2-rev5b-onegaminvs.nb for the one-\epsilon ones.
%
\begin{equation}
\bigg(\Big( D_{m} F_{m_1 m_2 m_3 m_4 m_5}^+\,
      D_{n} F_{n_1 n_2 n_3 n_4 n_5}^+ \Big)
 \Big|_{\widetilde{\tableau{2 1 1 1 1}}^+\!\otimes\;
        \widetilde{\tableau{2 1 1 1 1}}^+} \times \eta^{m n} \eta^{m_5 n_5}
\bigg)\bigg|_{\widetilde{\tableau{2 2 2 2}}}
 = 0\, .
\end{equation}
The other contraction does lead to a non-vanishing result:
% 
% W3F2-rev5b-twogamBr.nb, as well as 
% W3F2-rev5b-onegaminvs.nb.
%
\begin{equation}
\Big( D_{m} F_{m_1 m_2 m_3 m_4 m_5}^+\,
      D_{n} F_{n_1 n_2 n_3 n_4 n_5}^+ \Big)
 \Big|_{\widetilde{\tableau{2 1 1 1 1}}^+\!\otimes\;
        \widetilde{\tableau{2 1 1 1 1}}^+} \times \eta^{m n_5} \eta^{m_5 n}
\not=0\, .
\end{equation}
The explicit expression can easily be obtained from~\eqn{e:DF5explicit}.

The invariants used in the action~\eqn{e:finalresult} can be
constructed from the ingredients given above. In the following we will
suppress all terms proportional to $D^m F_{m r_2\cdots r_5}^+$ as they
are lowest-order equations of motion. Within each invariant, the terms
can be classified according to the number of indices that are
contracted between the $W^2$ and the $(DF_{(5)}^+)^2$ blocks. For space
reasons, we list here only the terms with the maximal number of
contractions between the two blocks.\footnote{Note, however, that the
lower-rank contractions arising from the trace-subtraction terms
in~\eqn{e:w21}--\eqn{e:w28} (or, alternatively, those in the corresponding
$(DF_{(5)}^+)^2$ expressions) are crucial to obtain a correct decomposition 
of the action in irreducible parts.}
These are given by
%
% W3F2-rev5b-finally.nb (the actual computation is elsewhere, this is
% where the summary is stored).
%
% w2dfdg8Brtrl:
% w2dfdg6Brtrl: 
% w2dfdg4Brtrl: 
%
\begin{align}
\label{e:w2dfdg8Brtrl}
W^2\Big|_{\widetilde{\tableau{2 2 2 2}}}
(DF_{(5)}^+\Big|_{\widetilde{\tableau{2 1 1 1 1}}^+})^2 &=
\begin{aligned}[t]
&  - \tfrac{5}{9} D_{k} F_{ d_1 d_2 d_3 d_4} \, D_{l} F_{ d_5 d_6 d_7 k} \, W_{d_1 d_2 d_7 l} W_{d_3 d_4 d_5 d_6}\\[1ex]
&  - \tfrac{5}{18} D_{l} F_{ d_1 d_2 d_3 d_4} \, D_{l} F_{ d_5 d_6 d_7 d_8} \, W_{d_1 d_2 d_5 d_6} W_{d_3 d_4 d_7 d_8}\\[1ex]
&  + \tfrac{5}{18} D_{k} F_{ d_1 d_2 d_3 d_4 l} \, D_{l} F_{ d_5 d_6 d_7 d_8 k} \, W_{d_1 d_2 d_5 d_6} W_{d_3 d_4 d_7 d_8}\\[1ex]
&  - \tfrac{5}{9} D_{k} F_{ d_1 d_2 d_3 l} \, D_{l} F_{ d_4 d_5 d_6 d_7} \, W_{d_1 d_2 d_6 d_7} W_{d_3 k d_4 d_5}\\[1ex]
&  + \tfrac{2}{9} D_{k} F_{ d_1 d_2 d_3} \, D_{l} F_{ d_4 d_5 d_6} \, W_{d_1 d_2 d_6 l} W_{d_3 k d_4 d_5}\\[1ex]
&  + \tfrac{2}{9} D_{k} F_{ d_1 d_2 d_3} \, D_{l} F_{ d_4 d_5 d_6} \, W_{d_1 d_2 d_4 d_5} W_{d_3 k d_6 l}\\[1ex]
&  - 2 D_{k} F_{ d_1 d_2 d_3} \, D_{l} F_{ d_4 d_5 d_6} \, W_{d_1 d_2 d_6 k} W_{d_3 l d_4 d_5}\\[1ex]
&  - 2 D_{k} F_{ d_1 d_2 d_3} \, D_{l} F_{ d_4 d_5 d_6} \, W_{d_1 d_2
	 d_4 d_5} W_{d_3 l d_6 k}\\[1ex]
& +\text{\small lower-rank contractions}\, ,
\end{aligned}
\end{align}
\begin{align}
%
% w2dfdg2x36Br:
% w2dfdg2x34Br:
%
\label{e:w2dfdg4BrtrlA}
W^2\Big|_{\widetilde{\tableau{2 2 2}}}
(DF_{(5)}^+\Big|_{\widetilde{\tableau{2 1 1 1 1}}^+})^2 &=
\begin{aligned}[t]
&    \tfrac{1}{6} D_{k} F_{ d_1 d_2} \, D_{l} F_{ d_3 d_4} \, W_{d_1 d_5 d_3 k} W_{d_2 d_4 d_5 l}\\[1ex]
&  - \tfrac{1}{36} D_{k} F_{ d_1 d_2} \, D_{l} F_{ d_3 d_4} \, W_{d_1 k d_5 l} W_{d_2 d_5 d_3 d_4}\\[1ex]
&  + \tfrac{1}{12} D_{k} F_{ d_1 d_2} \, D_{l} F_{ d_3 d_4} \, W_{d_1 l d_5 k} W_{d_2 d_5 d_3 d_4}\\[1ex]
&  - \tfrac{1}{18} D_{k} F_{ d_1 d_2} \, D_{l} F_{ d_3 d_4} \, W_{d_1 d_3 d_5 k} W_{d_2 d_5 d_4 l}\\[1ex]
&  - \tfrac{1}{9} D_{k} F_{ d_1 d_2} \, D_{l} F_{ d_3 d_4} \, W_{d_1 d_5 d_3 k} W_{d_2 d_5 d_4 l}\\[1ex]
&  - \tfrac{1}{72} D_{k} F_{ d_1 d_2} \, D_{l} F_{ d_3 d_4} \, W_{d_1 d_2 d_5 l} W_{d_3 d_4 d_5 k}\\[1ex]
&  + \tfrac{1}{24} D_{k} F_{ d_1 d_2} \, D_{l} F_{ d_3 d_4} \, W_{d_1 d_2 d_5 k} W_{d_3 d_4 d_5 l}\\[1ex]
&  + \tfrac{1}{6} D_{l} F_{ d_1 d_2 d_3} \, D_{l} F_{ d_4 d_5 d_6} \, W_{d_1 d_2 d_6 d_7} W_{d_3 d_7 d_4 d_5}\\[1ex]
&  + \tfrac{1}{6} D_{k} F_{ d_1 d_2 d_3 l} \, D_{l} F_{ d_4 d_5 d_6 k} \, W_{d_1 d_2 d_6 d_7} W_{d_3 d_7 d_4 d_5}\\[1ex]
&  + \tfrac{1}{12} D_{k} F_{ d_1 d_2} \, D_{l} F_{ d_3 d_4} \, W_{d_1 d_2 d_4 d_5} W_{d_3 k d_5 l}\\[1ex]
&  - \tfrac{1}{36} D_{k} F_{ d_1 d_2} \, D_{l} F_{ d_3 d_4} \, W_{d_1 d_2 d_4 d_5} W_{d_3 l d_5 k}\\[1ex]
& +\text{\small lower-rank contractions}\,,
\end{aligned}\\[1ex]
%
% w2dfdgB20011:
%
\label{e:w2dfdgB20011}
W^2\Big|_{\widetilde{\tableau{3 1 1 1}}}
(DF_{(5)}^+\Big|_{\widetilde{\tableau{2 1 1 1 1}}^+})^2 &=
\begin{aligned}[t]
&    \tfrac{1}{6} D_{k} F_{ d_1 d_2} \, D_{l} F_{ d_3 d_4} \, W_{d_1 d_5 d_3 k} W_{d_2 d_4 d_5 l}\\[1ex]
&  - \tfrac{1}{36} D_{k} F_{ d_1 d_2} \, D_{l} F_{ d_3 d_4} \, W_{d_1 k d_5 l} W_{d_2 d_5 d_3 d_4}\\[1ex]
&  + \tfrac{1}{12} D_{k} F_{ d_1 d_2} \, D_{l} F_{ d_3 d_4} \, W_{d_1 l d_5 k} W_{d_2 d_5 d_3 d_4}\\[1ex]
&  - \tfrac{1}{18} D_{k} F_{ d_1 d_2} \, D_{l} F_{ d_3 d_4} \, W_{d_1 d_3 d_5 k} W_{d_2 d_5 d_4 l}\\[1ex]
&  - \tfrac{1}{9} D_{k} F_{ d_1 d_2} \, D_{l} F_{ d_3 d_4} \, W_{d_1 d_5 d_3 k} W_{d_2 d_5 d_4 l}\\[1ex]
&  - \tfrac{1}{72} D_{k} F_{ d_1 d_2} \, D_{l} F_{ d_3 d_4} \, W_{d_1 d_2 d_5 l} W_{d_3 d_4 d_5 k}\\[1ex]
&  + \tfrac{1}{24} D_{k} F_{ d_1 d_2} \, D_{l} F_{ d_3 d_4} \, W_{d_1 d_2 d_5 k} W_{d_3 d_4 d_5 l}\\[1ex]
&  + \tfrac{1}{6} D_{l} F_{ d_1 d_2 d_3} \, D_{l} F_{ d_4 d_5 d_6} \, W_{d_1 d_2 d_6 d_7} W_{d_3 d_7 d_4 d_5}\\[1ex]
&  + \tfrac{1}{6} D_{k} F_{ d_1 d_2 d_3 l} \, D_{l} F_{ d_4 d_5 d_6 k} \, W_{d_1 d_2 d_6 d_7} W_{d_3 d_7 d_4 d_5}\\[1ex]
&  + \tfrac{1}{12} D_{k} F_{ d_1 d_2} \, D_{l} F_{ d_3 d_4} \, W_{d_1 d_2 d_4 d_5} W_{d_3 k d_5 l}\\[1ex]
&  - \tfrac{1}{36} D_{k} F_{ d_1 d_2} \, D_{l} F_{ d_3 d_4} \, W_{d_1 d_2 d_4 d_5} W_{d_3 l d_5 k}\\[1ex]
& +\text{\small lower-rank contractions}\,,
\end{aligned}\\[1ex]
%
% w2dfdgB2001102000Atrl:
%
\label{e:w2dfdgB02000A}
W^2\Big|_{\widetilde{\tableau{2 2}}_A}
(DF_{(5)}^+\Big|_{\widetilde{\tableau{2 1 1 1 1}}^+})^2 &=
\begin{aligned}[t]
&  - \tfrac{1}{72} D_{k} F_{ d_1 l} \, D_{l} F_{ d_2 d_3} \, W_{d_1 k d_4 d_5} W_{d_2 d_3 d_4 d_5}\\[1ex]
&  + \tfrac{1}{144} D_{k} F_{ d_1} \, D_{l} F_{ d_2} \, W_{d_1 d_4 d_3 k} W_{d_2 d_3 d_4 l}\\[1ex]
&  + \tfrac{1}{18} D_{k} F_{ d_1 d_2} \, D_{l} F_{ d_3 k} \, W_{d_1 d_5 d_4 l} W_{d_2 d_4 d_3 d_5}\\[1ex]
&  + \tfrac{1}{24} D_{k} F_{ d_1} \, D_{l} F_{ d_2} \, W_{d_1 d_3 d_4 l} W_{d_2 d_4 d_3 k}\\[1ex]
&  - \tfrac{7}{144} D_{k} F_{ d_1} \, D_{l} F_{ d_2} \, W_{d_1 d_4 d_3 l} W_{d_2 d_4 d_3 k}\\[1ex]
&  + \text{\small p.t.o.}
\end{aligned}
\end{align}
\begin{align}
&\phantom{=}\;  \begin{aligned}[t]
&  - \tfrac{7}{144} D_{k} F_{ d_1} \, D_{l} F_{ d_2} \, W_{d_1 d_4 d_3 k} W_{d_2 d_4 d_3 l}\\[1ex]
&  - \tfrac{1}{144} D_{l} F_{ d_1 d_2} \, D_{l} F_{ d_3 d_4} \, W_{d_1 d_3 d_5 d_6} W_{d_2 d_4 d_5 d_6}\\[1ex]
&  + \tfrac{1}{48} D_{k} F_{ d_1 d_2 l} \, D_{l} F_{ d_3 d_4 k} \, W_{d_1 d_3 d_5 d_6} W_{d_2 d_4 d_5 d_6}\\[1ex]
&  + \tfrac{1}{72} D_{l} F_{ d_1 d_2} \, D_{l} F_{ d_3 d_4} \, W_{d_1 d_5 d_3 d_6} W_{d_2 d_5 d_4 d_6}\\[1ex]
&  - \tfrac{1}{24} D_{k} F_{ d_1 d_2 l} \, D_{l} F_{ d_3 d_4 k} \, W_{d_1 d_5 d_3 d_6} W_{d_2 d_5 d_4 d_6}\\[1ex]
&  + \tfrac{1}{18} D_{k} F_{ d_1 l} \, D_{l} F_{ d_2 d_3} \, W_{d_1 d_4 d_3 d_5} W_{d_2 d_5 d_4 k}\\[1ex]
&  - \tfrac{1}{36} D_{k} F_{ d_1 l} \, D_{l} F_{ d_2 d_3} \, W_{d_1 d_3 d_4 d_5} W_{d_2 k d_4 d_5}\\[1ex]
&  - \tfrac{1}{288} D_{l} F_{ d_1 d_2} \, D_{l} F_{ d_3 d_4} \, W_{d_1 d_2 d_5 d_6} W_{d_3 d_4 d_5 d_6}\\[1ex]
&  + \tfrac{1}{96} D_{k} F_{ d_1 d_2 l} \, D_{l} F_{ d_3 d_4 k} \, W_{d_1 d_2 d_5 d_6} W_{d_3 d_4 d_5 d_6}\\[1ex]
&  - \tfrac{1}{288} D_{k} F_{ d_1} \, D_{l} F_{ d_2} \, W_{d_1 d_2 d_3 d_4} W_{d_3 d_4 k l}\\[1ex]
&  + \tfrac{7}{144} D_{k} F_{ d_1} \, D_{l} F_{ d_2} \, W_{d_1 d_3 d_2 d_4} W_{d_3 k d_4 l}\\[1ex]
&  - \tfrac{1}{72} D_{k} F_{ d_1 d_2} \, D_{l} F_{ d_3 k} \, W_{d_1 d_2 d_4 d_5} W_{d_3 l d_4 d_5}\\[1ex]
& +\text{\small lower-rank contractions}\,,
\end{aligned}\nonumber\\[1ex]
%
% w2dfdgB02000Strl:
%
\label{e:w2dfdgB02000S}
W^2\Big|_{\widetilde{\tableau{2 2}}_S}
(DF_{(5)}^+\Big|_{\widetilde{\tableau{2 1 1 1 1}}^+})^2 &=
\begin{aligned}[t]
&    \tfrac{1}{36} D_{k} F_{ d_1 d_2} \, D_{l} F_{ d_3 k} \, W_{d_1 l d_4 d_5} W_{d_2 d_3 d_4 d_5}\\[1ex]
&  + \tfrac{1}{24} D_{k} F_{ d_1} \, D_{l} F_{ d_2} \, W_{d_1 d_4 d_3 k} W_{d_2 d_3 d_4 l}\\[1ex]
&  - \tfrac{1}{9} D_{k} F_{ d_1 d_2} \, D_{l} F_{ d_3 k} \, W_{d_1 d_5 d_4 l} W_{d_2 d_4 d_3 d_5}\\[1ex]
&  + \tfrac{1}{144} D_{k} F_{ d_1} \, D_{l} F_{ d_2} \, W_{d_1 d_3 d_4 l} W_{d_2 d_4 d_3 k}\\[1ex]
&  + \tfrac{1}{144} D_{k} F_{ d_1} \, D_{l} F_{ d_2} \, W_{d_1 d_4 d_3 l} W_{d_2 d_4 d_3 k}\\[1ex]
&  + \tfrac{1}{24} D_{k} F_{ d_1} \, D_{l} F_{ d_2} \, W_{d_1 d_4 d_3 k} W_{d_2 d_4 d_3 l}\\[1ex]
&  + \tfrac{1}{144} D_{l} F_{ d_1 d_2} \, D_{l} F_{ d_3 d_4} \, W_{d_1 d_3 d_5 d_6} W_{d_2 d_4 d_5 d_6}\\[1ex]
&  - \tfrac{1}{48} D_{k} F_{ d_1 d_2 l} \, D_{l} F_{ d_3 d_4 k} \, W_{d_1 d_3 d_5 d_6} W_{d_2 d_4 d_5 d_6}\\[1ex]
&  - \tfrac{1}{36} D_{l} F_{ d_1 d_2} \, D_{l} F_{ d_3 d_4} \, W_{d_1 d_5 d_3 d_6} W_{d_2 d_5 d_4 d_6}\\[1ex]
&  + \tfrac{1}{12} D_{k} F_{ d_1 d_2 l} \, D_{l} F_{ d_3 d_4 k} \, W_{d_1 d_5 d_3 d_6} W_{d_2 d_5 d_4 d_6}\\[1ex]
&  - \tfrac{1}{9} D_{k} F_{ d_1 l} \, D_{l} F_{ d_2 d_3} \, W_{d_1 d_4 d_3 d_5} W_{d_2 d_5 d_4 k}\\[1ex]
&  + \tfrac{1}{36} D_{k} F_{ d_1 l} \, D_{l} F_{ d_2 d_3} \, W_{d_1 d_3 d_4 d_5} W_{d_2 k d_4 d_5}\\[1ex]
&  + \tfrac{7}{288} D_{k} F_{ d_1} \, D_{l} F_{ d_2} \, W_{d_1 d_2 d_3 d_4} W_{d_3 d_4 k l}\\[1ex]
&  - \tfrac{7}{72} D_{k} F_{ d_1} \, D_{l} F_{ d_2} \, W_{d_1 d_3 d_2 d_4} W_{d_3 k d_4 l}\\[1ex]
& +\text{\small lower-rank contractions}\,,
\end{aligned}
\end{align}
\begin{align}
%
% w2dfdgB40000:
%
\label{e:w2dfdgB40000}
W^2\Big|_{\widetilde{\tableau{4}}}
(DF_{(5)}^+\Big|_{\widetilde{\tableau{2 1 1 1 1}}^+})^2 &= 
\begin{aligned}[t]
&    \tfrac{1}{6} D_{k} F_{ d_1} \, D_{l} F_{ d_2} \, W_{d_1 d_4 d_3 k} W_{d_2 d_3 d_4 l}\\[1ex]
&  + \tfrac{1}{6} D_{k} F_{ d_1} \, D_{l} F_{ d_2} \, W_{d_1 d_3 d_4 l} W_{d_2 d_4 d_3 k}\\[1ex]
&  + \tfrac{1}{6} D_{k} F_{ d_1} \, D_{l} F_{ d_2} \, W_{d_1 d_4 d_3 l} W_{d_2 d_4 d_3 k}\\[1ex]
&  + \tfrac{1}{6} D_{k} F_{ d_1} \, D_{l} F_{ d_2} \, W_{d_1 d_4 d_3 k} W_{d_2 d_4 d_3 l}\\[1ex]
&  - \tfrac{1}{12} D_{k} F_{ d_1} \, D_{l} F_{ d_2} \, W_{d_1 d_2 d_3 d_4} W_{d_3 d_4 k l}\\[1ex]
&  + \tfrac{1}{3} D_{k} F_{ d_1} \, D_{l} F_{ d_2} \, W_{d_1 d_3 d_2 d_4} W_{d_3 k d_4 l}\\[1ex]
& +\text{\small lower-rank contractions}\,,
\end{aligned}\\[1ex]
%
% w2dfdg20000:
%
W^2\Big|_{\widetilde{\tableau{2}}}
(DF_{(5)}^+\Big|_{\widetilde{\tableau{2 1 1 1 1}}^+})^2 &= 
\begin{aligned}[t]
&    \tfrac{1}{144} D_{k} F_{} \, D_{l} F_{} \,\, W_{d_1 k d_2 d_3} W_{d_1 l d_2 d_3}\\[1ex]
&  + \tfrac{9}{144} D_{l} F_{ d_1} \, D_{l} F_{ d_2} \, W_{d_1 d_3 d_4 d_5} W_{d_2 d_3 d_4 d_5}\\[1ex]
&  + \tfrac{7}{36} D_{k} F_{ d_1 l} \, D_{l} F_{ d_2 k} \, W_{d_1 d_3 d_4 d_5} W_{d_2 d_3 d_4 d_5}\\[1ex]
&  - \tfrac{1}{144} D_{k} F_{ l} \, D_{l} F_{ d_1} \, W_{d_1 d_2 d_3 d_4} W_{d_2 k d_3 d_4}\\[1ex]
&  - \tfrac{1}{144} D_{k} F_{ d_1} \, D_{l} F_{ k} \, W_{d_1 d_2 d_3 d_4} W_{d_2 l d_3 d_4}\\[1ex]
& +\text{\small lower-rank contractions}\,.
\end{aligned}
\end{align}
In the above expressions we have dropped all index pairs contracted
between the two field strengths, so
$F_{d_1d_2d_3d_4}F_{d_5d_6d_7d_8}\equiv F_{d_1d_2d_3d_4}{}^m
F_{d_5d_6d_7d_8m}$, and so on. For ease of notation, we have also left 
out the labels indicating that $F_{(5)}$ is self-dual. 
As was the case for the amplitude trace, the $(DF_{(5)}^+)^2$ tensors
were obtained as the sum of parity-even and -odd parts after the
extraction of a self-duality projector 
$\mathcal{P}_+=\tfrac{1}{2}(\mathbb{1}+*)$ from each of the $F_{(5)}^+$
tensors.

In the actual computation, the invariants constructed from $DF_{(5)}^+$ 
in the fully antisymmetric representation are also required in order to 
complete the basis. This has enabled us to make a strong consistency 
check on the calculation, as it allowed us to verify that the amplitude 
can indeed be expanded in the basis of invariants predicted by group 
theory. Because such terms vanish identically on shell and therefore
can be left out of the effective action, and because they are rather 
lengthy expressions, we do not list them here.

\end{sectionunit}

\begin{sectionunit}
\title{Fermionic world-sheet correlators}
\maketitle
\label{a:fermcorr}
The general world-sheet correlator for fermions was derived
by~\dcite{Atick:1987rs}; it takes the form
\begin{multline}
\label{e:aticksenmaster}
\Big\langle \prod_{i=1}^{N_1} S^+(\widetilde y_i)
\prod_{i=1}^{N_2} S^-(y_i) \prod_{i=1}^{N_3} \bar\Psi(\widetilde z_i)
\prod_{i=1}^{N_4} \Psi(z_i)\Big\rangle_\nu \\[1ex]
= \begin{aligned}[t]
{} & K_\nu  \frac{\displaystyle \prod_{i<j}\theta_1(\widetilde y_i-\widetilde y_j)^{\tfrac{1}{4}}
\prod_{i<j} \theta_1(y_i-y_j)^{\tfrac{1}{4}} 
\prod_{i<j} \theta_1(\widetilde z_i-\widetilde z_j)
\prod_{i<j} \theta_1(z_i- z_j)}{\displaystyle
  \prod_{i,j} \theta_1(y_i-\widetilde y_j)^{\tfrac{1}{4}}
  \prod_{i,j} \theta_1(z_i-\widetilde z_i)
} \\[1ex]
& \quad\quad\times 
\frac{ \displaystyle
  \prod_{i,j} \theta_1(z_i-\widetilde y_j)^{\tfrac{1}{2}}
  \prod_{i,j} \theta_1(\widetilde z_i - y_j)^{\tfrac{1}{2}}}{\displaystyle
  \prod_{i,j} \theta_1(\widetilde z_i- \widetilde y_j)^{\tfrac{1}{2}}
  \prod_{i,j} \theta_1(z_i-y_j)^{\tfrac{1}{2}}
}\\[1ex]
&\quad\quad\times
\theta_\nu\left( \frac{1}{2}\sum_i \widetilde y_i - \frac{1}{2}\sum_i y_i 
+ \sum_i z_i - \sum_i \widetilde z_i\right)\, .
\end{aligned}
\end{multline}
The normalisation constant $K_\nu$ is determined by taking limits of
the insertion points such that all~$\theta_1$ functions reduce to
poles or zeroes, for which one uses
\begin{equation}
\label{e:reduceth1}
\lim_{z\rightarrow 0} \theta_1(z) = \theta_1'(0) \cdot z\, .
\end{equation}
Because of the operator product expansion, the total correlator in this
limit should be equal to the product of these poles and zeroes, times 
the expectation value of the identity in each spin structure,
\begin{equation}
\langle 1 \rangle_\nu =
\left(\frac{\theta_\nu(0)}{\theta_1'(0)}\right)^4\, .
\end{equation}
In the situations we analyse, the $\theta_\nu$ factors of the
fermionic correlator and the one from the ghost correlator conspire to
give the $\theta_\nu(0)$. Therefore, the procedure described here
fixes $K_\nu$ in terms of a power of $\theta_1'(0)$. In order to
determine the overall sign, one has keep track of the ordering of the
fermions, especially when converting the covariant expression to one
written in terms of the five helicity basis fermions.

As an example, consider the fifth line of table~\ref{t:twotwo}. The relevant fermion
correlator (multiplied with the one for the ghosts) is
\begin{multline}
\sum_{\nu} (-)^{\nu-1} \Big\langle S_{\alpha_1}(z_1) S_{\alpha_2}(z_2) 
     \Psi^1(z_3)\Psi^3(z_3) \Psi^3(z_4) \Psi^5(z_4) \Psi^5(v)\Big\rangle_\nu\; {\cal G}(z_1,z_2,v)_\nu\\
\begin{aligned}
= \smash{\sum_{\nu} (-)^{\nu-1}} &\Big\langle S_-(z_1) S_-(z_2) \big(\psi(z_3) + \bar{\psi}(z_3)\big) \Big\rangle_\nu \\
  &\Big\langle S_+(z_1) S_-(z_2) \big(\psi(z_3) + \bar{\psi}(z_3)\big)\big(\psi(z_4) + \bar{\psi}(z_4)\big) \Big\rangle_\nu \\
  &\Big\langle S_+(z_1) S_-(z_2) \big(\psi(z_4) + \bar{\psi}(z_4)\big)\big(\psi(v) + \bar{\psi}(v)\big) \Big\rangle_\nu\, \\
  &\Big\langle S_+(z_1) S_-(z_2) \Big\rangle_\nu^2\,{\cal G}(z_1,z_2,v)_\nu\,.
\end{aligned}
\end{multline}
In the limit $z_1\rightarrow z_2$ and $z_3\rightarrow z_4$ one can
use~\eqn{e:reduceth1} to simplify the result obtained
using~\eqn{e:aticksenmaster}. One can also analyse the expected
behaviour of the amplitude by looking at the operator product expansion. Using the ghost correlator
\begin{equation}
{\cal G}(z_1,z_2,v)_\nu = \theta_1(v-z_1)^{\tfrac{1}{2}}\, 
                          \theta_1(v-z_2)^{\tfrac{1}{2}}\,
                          \theta_1(z_2-z_1)^{-\tfrac{1}{4}}\,
                          \theta_\nu\left(\frac{z_1-z_2}{2}\right)^{-1}\, ,
\end{equation}
one finds the limiting expression
\begin{multline}
\sum_\nu (-)^{\nu-1} \left(\frac{\theta_\nu(0)}{\theta'_1(0)}\right)^4
\frac{4\,(v-z_2)(z_2-z_3)^2}{(z_3-v)(z_1-z_3)(z_3-z_4)(z_2-z_3)(z_1-z_3)(z_1-z_2)}\\[1ex]
\times
\begin{cases}
-K_\nu\, \theta'_1(0)  & \text{from~\eqn{e:aticksenmaster}\,,} \\
1             & \text{from the OPEs\,.}
\end{cases}
\end{multline}
Comparing the two, one thus deduces that the normalisation constant is
$K_\nu = -1/\theta'_1(0)$ (independent of the spin structure sector). 

Having determined the normalisation, the sum over spin structures is then evaluated
using the Riemann identity
\begin{multline}
\label{e:Riemann_id}
\sum_{\nu=1,2,3,4} (-)^{\nu-1}\, \theta_\nu(z_1|\tau) \theta_\nu(z_2|\tau)
\theta_\nu(z_3|\tau) \theta_\nu(z_4|\tau) \\[1ex] = 
2\,
\theta_1\left(\frac{z_1+z_2+z_3+z_4}{2}\Big|\tau\right)
\theta_1\left(\frac{z_1+z_2-z_3-z_4}{2}\Big|\tau\right)\\[1ex]
\times\theta_1\left(\frac{z_1-z_2-z_3+z_4}{2}\Big|\tau\right)
\theta_1\left(\frac{z_1-z_2+z_3-z_4}{2}\Big|\tau\right)\, .
\end{multline}

\end{sectionunit}
\end{sectionunit}

\vfill\eject
%\bibliography{kasbib}
\bibliography{fourpoint}
\end{document}